\documentclass[superscriptaddress,reprint,amsmath,amssymb,aps, prb]{revtex4-1}
\usepackage{graphicx}
\usepackage{dcolumn}
\usepackage{bm}
\usepackage{braket}
\usepackage{mathtools}
\usepackage{amsfonts}
\usepackage[breaklinks,colorlinks=true,linkcolor=red,urlcolor=blue,citecolor=blue]{hyperref}

{\par\noindent{\em #1:\ }}%
{~\rule{2mm}{2mm}\par\bigskip}

\begin{document}
\title{Oriented propagation of magnetization due to chiral edge modes\\ in Kitaev-type models}

\author{Tomonari Mizoguchi}
\email{mizoguchi@rhodia.ph.tsukuba.ac.jp}
\affiliation{Department of Physics, University of Tsukuba, Tsukuba, Ibaraki 305-8571, Japan}
\author{Tohru Koma}
\email{tohru.koma@gakushuin.ac.jp}
\affiliation{Department of Physics, Gakushuin University, Mejiro, Toshima-ku, Tokyo 171-8588, Japan}
\author{Yasuhito Yoshida}
\affiliation{Department of Engineering Science, University of Electro-Communications, Chofu, Tokyo 182-8585, Japan}

\date{\today}
\begin{abstract}
Detecting chiral edge modes in topological materials has been
intensively pursued in experiments.
However, the phenomena caused by the modes are not yet elucidated
theoretically.
We study the dynamics of chiral spinon wave packets at the edge
in Kitaev-type magnets. More precisely, by relying on the exact
solvability of the models, we construct a spinon wave packet,
localized edge magnetization, which shows oriented propagation
along the edge, whose behavior is expected from the chiral character
of the dispersion relation of the chiral edge modes.
In general, this approach enables us to study not only spin transport
in anisotropic magnets but also charge transport in Bogoliubov-de Gennes-type
superconductors because it does not rely on a conserved quantity.
\end{abstract}

\maketitle
\section{Introduction \label{sec:intro}}
The exact ground state of the Kitaev honeycomb model~\cite{Kitaev2006} is known as 
a quantum spin liquid~\cite{Anderson1973,Balents2010,Zhou2017,Knolle2019} that shows short-range spin-spin correlations~\cite{Baskaran2007}. 
As it is the first example of the exactly solvable model of a quantum spin liquid in two dimensions, 
tremendous theoretical studies have been done to reveal fundamental properties of this model such as thermal quantities and spin dynamics~\cite{Knolle2014,Nasu2014,Nasu2015,Knolle2015,Yoshitake2017,Yoshitake2017_2,Nasu2018,Gohlke2018,Motome2019}. 
In parallel with theoretical studies, experimental attempts to seek the Kitaev spin liquids in 
candidate materials~\cite{Jackeli2009,Singh2010,Singh2012,Plumb2014,Rau2014,Trebst2017}
are ongoing~\cite{Sandilands2015,Banerjee2016,Do2017,Banerjee2017,Winter2017,Hermanns2018}.

In addition to the nature of the quantum spin liquid, the model is a weak topological material 
that hosts a Majorana edge flat band~\cite{Thakurathi2014} which yields unidirectional 
edge magnetization~\cite{Mizoguchi2019}. 
When applying an external 
magnetic field along [111] direction, the ground state of the model 
is believed to change to a chiral spin liquid which is characterized by 
a Chern number~\cite{Kitaev2006}, and exhibits chiral edge modes 
although the exact solvability is lost. 
If three spin interactions are added to the Hamiltonian of the Kitaev 
honeycomb model instead of the external magnetic field, then 
the exact solvability is retained, and the resulting model shows 
the character of a chiral spin liquid. A different variant of the model 
in the same topological class can be realized as an exactly solvable 
Kitaev-type model on a triangle-honeycomb lattice~\cite{Yao2007,Chung2010}. 
Thus, several Kitaev-type models were found to have the character of a chiral spin liquid. 

Detecting chiral edge modes~\cite{Katsura2010,Diaz2019} in topological materials has been 
a central issue in experiments as well. 
Actually, Kitaev-type magnets~\cite{Singh2010,Singh2012,Plumb2014} in the external magnetic field are most likely candidates 
whose ground state is a chiral spin liquid~\cite{Kasahara2018,Kasahara2018_2}. 
Therefore, a much deeper theoretical understanding of 
the related phenomena is required. 
In this paper, we study propagation of edge magnetization due to 
the chiral edge modes in Kitaev-type magnets. 
More precisely, by relying on the exact solvability of the models, 
we construct a spinon wave packet, localized edge magnetization, 
which is an excited state composed of the chiral edge states.   
Because of the chirality of the dispersion relation of the chiral edge mode, 
the wave packet propagates in only one of the two directions of the edge 
irrespectively of the initial form of the wave packet. 
In comparison with previous studies addressing spin transport which 
use spin current~\cite{Chen2013,Chatterjee2015,Hirobe2017,Carvalho2018}, the advantage of our method is that 
we need neither to introduce spin current nor to apply an external driving force 
such as a potential difference. Thus, our method can be applied to 
not only anisotropic magnets in which non of the spin components commutes with 
the Hamiltonian but also Bogoliubov-de Gennes (BdG)-type superconductors~\cite{Sato2016}
in which the charge is not conserved. 
We can expect that such oriented propagation of a local excitation at a sample edge 
is experimentally detectable. 

For the purpose of actually calculating the chiral edge modes, 
we also introduce a new technique into the transfer matrix method.
Although the method is already established as a standard tool~\cite{Lee1981,Hatsugai1993,Hatsugai1993_2,Dwivedi2016}, 
known exact solutions of chiral edge states are still rare so far;
the examples include
the Bernevig-Hughes-Zhang model~\cite{Mao2010,Dwivedi2016}
and the modified Haldane and Kane-Mele models~\cite{Kunst2019}.
Even in the case of the well-known Haldane tight-binding model 
on the honeycomb lattice~\cite{Haldane1988,Huang2012,Doh2014,Pantaleon2017,Kunst2019}, 
the exact solution of the chiral edge state has not yet been obtained.
In this paper, we demonstrate that our new technique enables us to obtain exact solutions of 
edge states, 
which is applicable to generic models described by $4 \times 4$ transfer matrices,  
including the Haldane model and the Kitaev model with three-spin interaction which we will deal with in this paper.

\begin{figure}[b]
\begin{center}
\includegraphics[width= 0.95\linewidth]{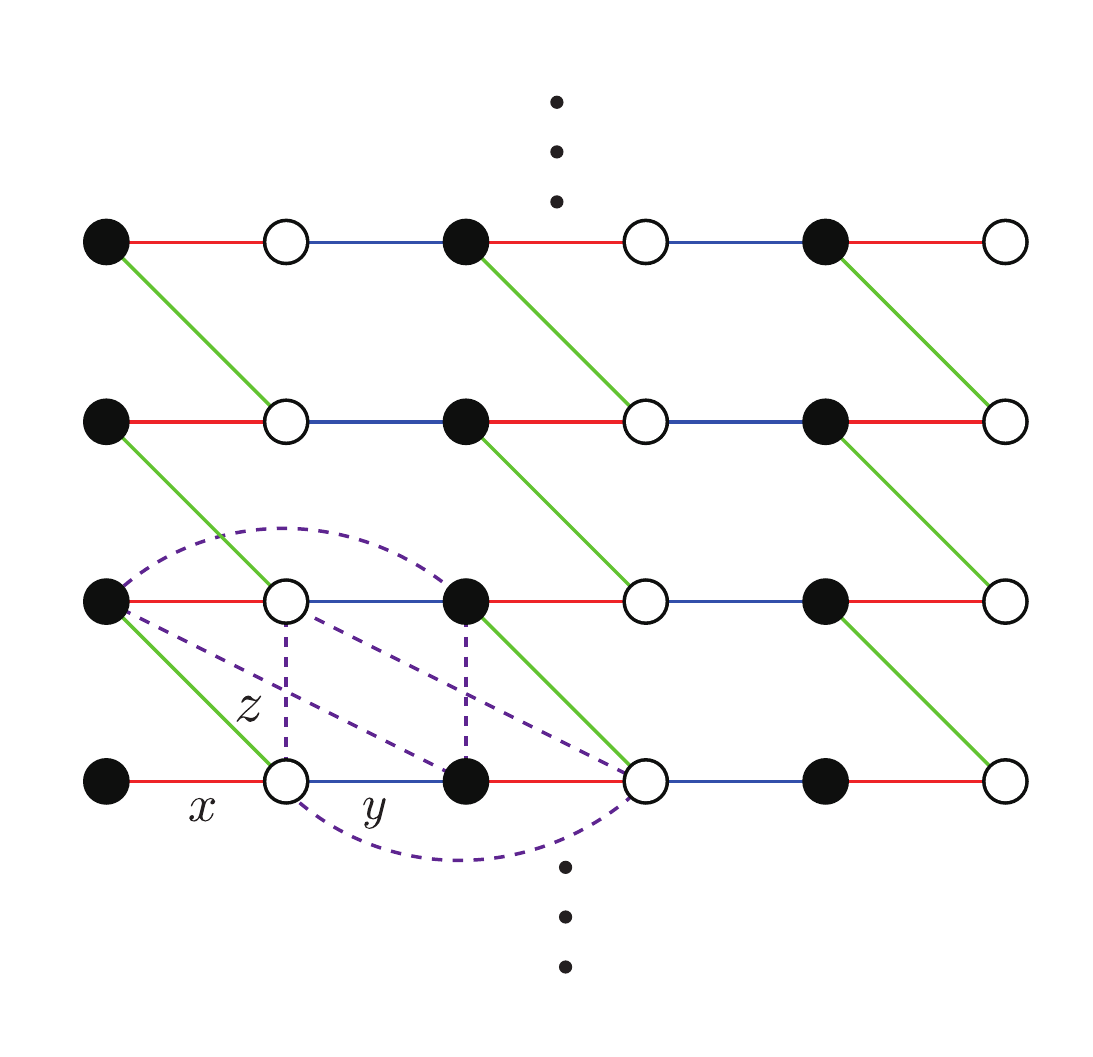}
\caption{The Kitaev's honeycomb model with edges in the horizontal direction.
Red, blue, and green bonds represent 
$x$- $y$- and $z$-bonds, respectively. 
Black (white) circles denote 
the sites with an odd (even) $\ell$. 
Purple dashed lines represent the next-nearest-neighbor hoppings 
of $c$-Majorana fermions induced by the three-spin interaction.}
\label{Fig1}
\end{center}
\end{figure}
The rest of this paper is organized as follows.
In Sec.~\ref{sec:model}, we describe the models considered in this paper, namely 
the Kitaev honeycomb model with three-spin interaction and the Kitaev triangle-honeycomb model.
We also explain the formalism of the Jordan-Wigner transformation 
by which these quantum spin models are mapped
to the fermion models whose Hamiltonian have BdG form which is interpreted as the Hamiltonian of a spinless superconductor. 
In Sec.~\ref{sec:chiraledge_transfer}, 
we derive the exact dispersion relation and the wave function of the chiral edge spinons 
by using the transfer matrix method. 
In  Sec.~\ref{sec:edgemag_propagate}, 
by relying on the exact solution thus obtained, we construct 
an excited state which realizes the oriented propagation of the edge magnetization. 
We also demonstrate that a wave packet constructed by this formulation indeed 
moves in only one of the two directions of the edge. 
In Sec.~\ref{sec:edge_DS}, the edge dynamical spin susceptibility is obtained as well. This quantity reflects the chiral nature of the edge mode,
and is experimentally detectable.
A summary and discussion are given in Sec.~\ref{sec:summary}. 
Appendix~\ref{sec:edgemode_method} is devoted to the details of the transfer matrix method used in Sec.~\ref{sec:chiraledge_transfer}.
Finally, in Appendix~\ref{sec:trihoney}, we present the details of the derivation 
of the exact solution of the chiral edge modes in the Kitaev triangle-honeycomb model. 

\section{Model \label{sec:model}}
In order to demonstrate that chiral edge modes in topological materials 
cause oriented propagation of local edge magnetization in Kitaev-type 
magnets, 
we treat two models, namely, the Kitaev honeycomb model with three-spin interactions, 
and the Kitaev triangle-honeycomb model.
In this section, we rewrite these Hamiltonians into the BdG form
which is interpreted as the Hamiltonian of a spinless superconductor.
If a reader is familiar with the following formulations, it is possible to skip this section.   
Of course, in order to understand the subsequent calculations, it is necessary to 
keep the notations in this section in mind.

\subsection{Kitaev honeycomb model with three-spin interaction}
We consider the following Hamiltonian on a cylindrical geometry, whose sites are 
denoted by $(\ell,m)$
with $\ell = 1, \cdots, 2L_x$ and $m=1, \cdots, L_y$ (Fig.~\ref{Fig1}):
\begin{equation}
H = H_{\mathrm{K}} +H_{3\mathrm{-spin}}, \label{eq:hamiltonian}
\end{equation}
where 
\begin{widetext}
\begin{eqnarray}
H_{\mathrm{K}} =  J_x \sum_{\ell = 1}^{L_x} \sum_{m = 1}^{L_y} \sigma^x_{(2\ell -1,m)} \sigma^x_{(2\ell,m)} 
 +  J_y\sum_{\ell = 1}^{L_x-1} \sum_{m = 1}^{L_y} \sigma^y_{(2\ell,m)} \sigma^y_{(2\ell  +1,m)}
+ J_z \sum_{\ell = 1}^{L_x} \sum_{m = 1}^{L_y} \sigma^z_{(2\ell,m)} \sigma^z_{(2\ell  -1,m+1)}, 
\label{eq:HamiltonianK}
\end{eqnarray}
and 
\begin{eqnarray}
H_{3 \mathrm{-spin}} &=&
J^\prime \sum_{\ell = 1}^{L_x} \sum_{m = 1}^{L_y} \left[ \sigma^x_{(2\ell -1,m)} \sigma^y_{(2\ell,m)} \sigma^z_{(2\ell-1,m+1)} +   \sigma^x_{(2\ell ,m+1)} \sigma^y_{(2\ell-1,m+1)} \sigma^z_{(2\ell,m)} \right] \nonumber \\
 &+& J^\prime \sum_{\ell = 1}^{L_x-1} \sum_{m = 1}^{L_y} \left[ \sigma^x_{(2\ell-1,m)} \sigma^z_{(2\ell,m)} \sigma^y_{(2\ell  +1,m)} +  \sigma^y_{(2\ell,m)} \sigma^z_{(2\ell+1,m)} \sigma^x_{(2\ell  +2,m)} \right] \nonumber \\
&+&J^\prime \sum_{\ell = 1}^{L_x-1} \sum_{m = 1}^{L_y} \left[ \sigma^z_{(2\ell-1 ,m+1 )} \sigma^x_{(2\ell,m)} \sigma^y_{(2\ell  +1,m)} + \sigma^y_{(2\ell ,m+1 )} \sigma^x_{(2\ell+1,m+1)} \sigma^z_{(2\ell  +2,m)} \right].
\label{eq:ham3spin}
\end{eqnarray}
\end{widetext}
We impose the periodic boundary condition,
$\bm{\sigma}_{(\ell, m+L_y)}=\bm{\sigma}_{(\ell,m)}$,
in the vertical direction,
while we impose the open boundary condition in the horizontal direction.  
Here, $H_{\mathrm{K}}$ represents the Kitaev honeycomb model~\cite{Kitaev2006},
where the real parameters, $J_x$, $J_y$, and $J_z$, are the exchange coupling on $x$, $y$, and $z$ bonds, respectively.
As to the classification of the bonds into three types, see Fig.~\ref{Fig1} and its caption. 
The Hamiltonian $H_{3 \mathrm{-spin}}$ represents the three-spin interaction with real coupling $J^\prime$. 
Although this term looks fairly artificial, 
it was shown by Kitaev that 
the interaction Hamiltonian
$H_{3 \mathrm{-spin}}$ naturally 
appears when the Zeeman energy of the external magnetic field in the [111] direction is treated within 
the third-order perturbation in the Kitaev honeycomb model~\cite{Kitaev2006}.
In the case with $J_x = J_y =J_z = J$,
the parameter $J^\prime$ can be estimated as
$J^\prime \sim \frac{h^3}{J^2}$,
where $h$ is the strength of the magnetic field.
Therefore, the Kitaev honeycomb model with the Zeeman term is believed to belong to 
the same topological class of the chiral spin liquid as that of the Kitaev honeycomb model 
with the three-spin interactions. 
This motivates us to study the chiral edge mode in the Kitaev honeycomb model 
with the three-spin interactions. 

To obtain the exact ground state of the Hamiltonian of Eq. (\ref{eq:hamiltonian}), 
we map the spin model to the fermion model, by using the Jordan-Wigner transformation.
Following Refs.~\onlinecite{Feng2007,Yao2007,Chen2007,Chen2008}, 
we introduce a complex fermion, $a_{(\ell,m)}$, and rewrite the spin operators such that 
\begin{subequations}
\begin{equation}
\sigma_{(\ell,m)}^+ = 2 a_{(\ell,m)}e^{i\pi\hat{\theta}_{(\ell,m)}}, \label{eq:JW_sigmax}
\end{equation}
\begin{equation}
\sigma_{(\ell,m)}^- = 2e^{i\pi\hat{\theta}_{(\ell,m)}} a^\dagger_{(\ell,m)}, \label{eq:JW_sigmay}
\end{equation}
\begin{equation}
\sigma_{(\ell,m)}^z = (-1)^\ell\left[ 2a_{(\ell,m)}^{\dagger}a_{(\ell,m)} -1 \right], \label{eq:JW_sigmaz}
\end{equation}
\end{subequations}
where $\sigma_i^{\pm} = \sigma_i^x \pm i\sigma_i^y$, and 
\begin{eqnarray}
\hat{\theta}_{(\ell,m)} =\sum_{m^\prime < m }\sum_{\ell^\prime = 1}^{2L_x} 
a^{\dagger}_{(\ell^\prime,m^\prime)}a_{(\ell^\prime,m^\prime)} 
+ \sum_{\ell^\prime < \ell }a^{\dagger}_{(\ell^\prime,m)}a_{(\ell^\prime,m)}. \nonumber \\
\end{eqnarray}
Further, we decompose the complex fermion $a_{(\ell,m)}$ into real and imaginary parts,
i.e., we introduce two Majorana fermions 
in the following manner~\cite{Feng2007,Yao2007,Chen2007,Chen2008}: 
\begin{eqnarray}
c_{(\ell,m)} = &  i \left[a_{(\ell,m)}^\dagger -a_{(\ell,m)}\right], \nonumber\\
d_{(\ell,m)} = & a_{(\ell,m)}^\dagger + a_{(\ell,m)}\ \ \mbox{if $\ell$ is odd}, \label{eq:Majoranaodd}
\end{eqnarray}
and 
\begin{eqnarray}
c_{(\ell,m)} = &a_{(\ell,m)}^\dagger + a_{(\ell,m)}, \nonumber\\
d_{(\ell,m)} =  & i\left[a_{(\ell,m)}^\dagger - a_{(\ell,m)}\right]\ \ \mbox{if $\ell$ is even}. \label{eq:Majoranaeven}
\end{eqnarray}
Then, by using $c$ and $d$, we can rewrite $ H_{\mathrm{K}}$ and $H_{3\mathrm{-spin}} $ as 
\begin{widetext}
\begin{align}
 H_{\mathrm{K}} =   i J_x \sum_{\ell = 1}^{L_x} \sum_{m = 1}^{L_y} c_{(2\ell -1,m)} c_{(2\ell,m)} 
 +   i J_y\sum_{\ell = 1}^{L_x-1} \sum_{m = 1}^{L_y} c_{(2\ell ,m)} c_{(2\ell + 1,m)}  +  
 J_z \sum_{\ell = 1}^{L_x} \sum_{m = 1}^{L_y} c_{(2\ell,m)} c_{(2\ell-1,m+1)}  d_{(2\ell,m)} d_{(2\ell-1,m+1)}, \label{eq:hami1d}
 \end{align}
and
\begin{eqnarray}
H_{3\mathrm{-spin}} &= &
J^\prime \sum_{\ell = 1}^{L_x} \sum_{m = 1}^{L_y} \left[ c_{(2\ell-1,m)}c_{(2\ell-1,m+1)}d_{(2\ell,m)}d_{(2\ell-1,m)} -  c_{(2\ell,m+1)}c_{(2\ell,m) } d_{(2\ell-1,m+1)}d_{(2\ell,m)} \right] \nonumber \\
 &+& i J^\prime \sum_{\ell = 1}^{L_x-1} \sum_{m = 1}^{L_y} \left[ c_{(2\ell+1,m)} c_{(2\ell-1,m)}  +  c_{(2\ell,m)} c_{(2\ell+2,m)}\right] \nonumber \\
&+ & J^\prime \sum_{\ell = 1}^{L_x-1} \sum_{m = 1}^{L_y} \left[ c_{(2\ell-1,m+1)}c_{(2\ell+1,m)}d_{(2\ell-1,m+1)}d_{(2\ell,m)}  + c_{(2\ell+2,m)}c_{(2\ell,m+1)} d_{(2\ell+1,m+1)} d_{(2\ell+2, m)} \right].
\label{eq:hami2d}
\end{eqnarray}
\end{widetext}
One finds that the pairs of $d_i d_j$ in Eqs.(\ref{eq:hami1d}) and (\ref{eq:hami2d}) commute with
$ H_{\mathrm{K}}$ and $H_{3\mathrm{-spin}}$, thus they are the conserved quantities.
Consequently, we can replace $d_i d_j$ with the classical number, as $d_i d_j = \pm i$. 
In the following, we call $d_i d_j$ a link variable.
In the ground state, the configuration of the link variables is determined so that the energy for the Hamiltonian for $c$-Majorana fermions is minimized; 
this can be achieved by 
$ d_{(2\ell,m)} d_{(2\ell-1,m+1)} = +i$ for $\ell = 1, \cdots, L_x$ and $m=1, \cdots, L_y$~\cite{Kitaev2006,Yao2007}.
Then, the Hamiltonian thus obtained can be written in the quadratic form of the $c$-Majorana fermions,  
\begin{widetext}
\begin{align}
 H_{\mathrm{K}} =   i J_x \sum_{\ell = 1}^{L_x} \sum_{m = 1}^{L_y} c_{(2\ell -1,m)} c_{(2\ell,m)} 
 +   i J_y\sum_{\ell = 1}^{L_x-1} \sum_{m = 1}^{L_y} c_{(2\ell ,m)} c_{(2\ell + 1,m)}  +  i J_z \sum_{\ell = 1}^{L_x} \sum_{m = 1}^{L_y} c_{(2\ell,m)} c_{(2\ell-1,m+1)}, \label{eq:hami1}
 \end{align}
and
\begin{eqnarray}
H_{3\mathrm{-spin}}&=&
i J^\prime \sum_{\ell = 1}^{L_x} \sum_{m = 1}^{L_y} \left[ c_{(2\ell-1,m)}c_{(2\ell-1,m+1)} +  c_{(2\ell,m+1)}c_{(2\ell,m) }\right] \nonumber \\
 &+& i J^\prime \sum_{\ell = 1}^{L_x-1} \sum_{m = 1}^{L_y} \left[ c_{(2\ell+1,m)} c_{(2\ell-1,m)} +  c_{(2\ell,m)} c_{(2\ell+2,m)}\right] \nonumber \\
&-&i J^\prime \sum_{\ell = 1}^{L_x-1} \sum_{m = 1}^{L_y} \left[ c_{(2\ell-1,m+1)}c_{(2\ell+1,m)}  + c_{(2\ell+2,m)}c_{(2\ell,m+1)}\right].
\label{eq:hami2}
\end{eqnarray}
\end{widetext}

To rewrite the Hamiltonian of Eqs. (\ref{eq:hami1}) and (\ref{eq:hami2}) in the BdG form, we introduce a complex fermion~\cite{Mizoguchi2019},
\begin{eqnarray} 
\alpha_{(\ell,m)} = \frac{c_{(2\ell-1,m)} + ic_{(2\ell,m)}}{2},
\end{eqnarray}
and perform the Fourier transformation in the vertical direction,
\begin{eqnarray}
\alpha_{\ell,k_y} = \frac{1}{\sqrt{L_y}}\sum_{m=1}^{L_y} e^{i k_y m} \alpha_{(\ell,m)}.
\end{eqnarray}
Using $\alpha_{\ell,k_y}$, we can write the Hamiltonian as
\begin{eqnarray}
H =  \sum_{k_y} \sum_{\ell, \ell^\prime} \bm{\Psi}^{\dagger}_\ell (k_y) [\hat{h}(k_y)]_{\ell,\ell^\prime}  \bm{\Psi}_{\ell^\prime} (k_y),  \label{eq:BDG}
\end{eqnarray}
with 
\begin{eqnarray}
\bm{\Psi}_\ell (k_y) = \left(\alpha_{\ell, k_y}, \alpha^\dagger_{\ell,-k_y} \right)^{\mathrm{T}},
\end{eqnarray}
and 
\begin{eqnarray}
 [\hat{h} (k_y)]_{\ell,\ell^\prime}  = 
 \left(
 \begin{array}{cc}
[h_0(k_y)]_{\ell, \ell^{\prime} } & [\Delta(k_y)]_{\ell,\ell^{\prime} } \\
\left[\Delta^{\dagger}(k_y) \right]_{\ell, \ell^{\prime} }  & [- h_0 (- k_y)]_{\ell, \ell^{\prime}} \\
\end{array}
\right).
\end{eqnarray}
$h_0(k_y)$, $\Delta(k_y)$, and $\Delta^{\dagger}(k_y)$ are defined as
\begin{eqnarray}
[h_0(k_y)]_{\ell, \ell^{\prime} } = \left(J_x -J_z \cos k_y \right) \delta_{\ell, \ell^\prime} -\frac{J_y}{2} \left( \delta_{\ell, \ell^\prime +1} + \delta_{\ell, \ell^\prime - 1} \right), \nonumber \\
\end{eqnarray}
\begin{eqnarray}
[\Delta(k_y)]_{\ell, \ell^{\prime} } &=&  
-i J_z \sin k_y  \nonumber \\
&-& 2 J^\prime \sin k_y - \left[ \frac{J_y}{2} + i J^\prime \left(e^{-i k_y} + 1\right) \right] 
\delta_{\ell, \ell^\prime + 1} \nonumber \\
&+& \left[ \frac{J_y}{2} + i J^\prime \left(e^{i k_y} + 1\right) \right] 
\delta_{\ell, \ell^\prime - 1}, \nonumber \\
\end{eqnarray}
and $\left[ \Delta^{\dagger}(k_y) \right]_{\ell, \ell^\prime} = \{ [\Delta(k_y)]_{\ell^\prime, \ell} \}^{\ast}$.

The topological class of the BdG Hamiltonian of Eq. (\ref{eq:BDG}) can be determined by the symmetries. 
For $J^\prime = 0$, it belongs to the BDI class, which has time-reversal, particle-hole, and chiral symmetries~\cite{Mizoguchi2019}.
As a result, the Majorana edge flat band appears in the A$_{\rm y}$ phase~\cite{Kitaev2006,Mizoguchi2019,Thakurathi2014}, 
which is protected by the weak topological nature. 
If $J^\prime$ is finite, then the particle-hole symmetry is kept, but the time-reversal symmetry is broken. 
Consequently, the chiral symmetry given by the product of the time-reversal and particle-hole symmetries is broken, too. 
Therefore, the corresponding topological class is D class, whose topological number is given by 
$\mathbb{Z}$, or the Chern number~\cite{Kitaev2006}. 
For this class, the chiral edge mode appears if the Chern number is nonzero.

\subsection{Kitaev triangle-honeycomb model}
Next, we describe the Kitaev triangle-honeycomb model~\cite{Yao2007},
which is another example of the Kitaev-type model having the chiral edge modes.  
\begin{figure}[b]
\begin{center}
\includegraphics[width=\linewidth]{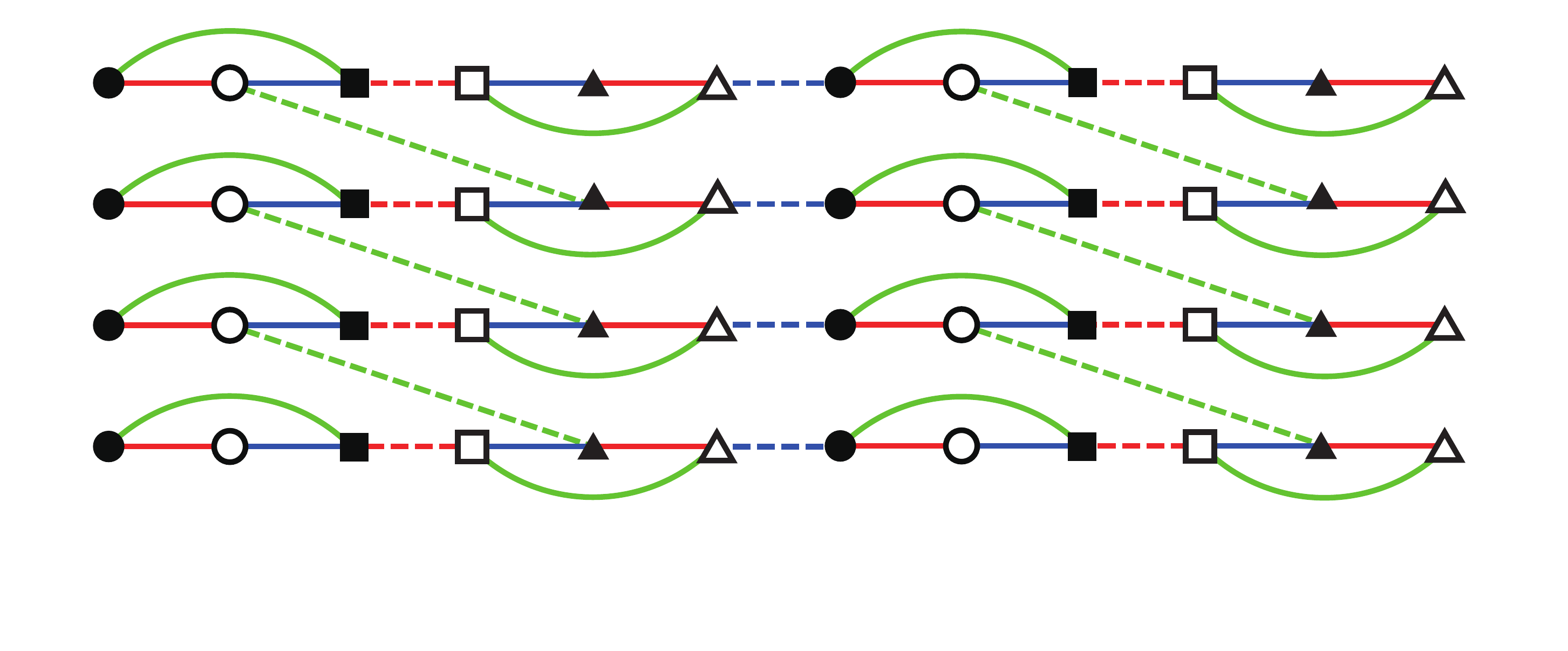}
\caption{The Kitaev's model on a triangle-honeycomb lattice.
On the red, blue, and green solid bonds, 
the exchange interactions are written as 
$J_x \sigma^x_{i} \sigma^x_j$, $J_y \sigma^y_{i} \sigma^y_j$, and $J_z \sigma^z_{i} \sigma^z_j$, respectively. 
On the red, blue, and green dashed bonds, 
the exchange interactions are written as 
$J^\prime_x \sigma^x_{i} \sigma^x_j$, $J^\prime_y \sigma^y_{i} \sigma^y_j$, and $J^\prime_z \sigma^z_{i} \sigma^z_j$, respectively. 
Note that $i,j$ are the abbreviation for the positions of sites. 
}
\label{Fig:TH}
\end{center}
\end{figure}
We consider a system with $6L_x \times L_y$ sites, and impose a periodic boundary condition, 
$\bm{\sigma}_{(\ell,m+L_y)} = \bm{\sigma}_{(\ell,m)}$, in the vertical direction.
The Kitaev triangle-honeycomb model is given as 
\begin{align}
H =& J_x \sum_{\ell = 1}^{L_x} \sum_{m = 1}^{L_y}[ \sigma^x_{(6\ell -5,m)} \sigma^x_{(6\ell-4 ,m)} + \sigma^x_{(6\ell -1,m)} \sigma^x_{(6\ell ,m)}  ] \notag \\
+& J_y\sum_{\ell = 1}^{L_x} \sum_{m = 1}^{L_y} [\sigma^y_{(6\ell -4,m)} \sigma^y_{(6\ell-3 ,m)} + \sigma^y_{(6\ell -2,m)} \sigma^y_{(6\ell-1 ,m)} ] \notag \\
+& J_z \sum_{\ell = 1}^{L_x} \sum_{m = 1}^{L_y}  [ \sigma^z_{(6\ell -5,m)} \sigma^z_{(6\ell-3 ,m)} + \sigma^z_{(6\ell -2,m)} \sigma^z_{(6\ell ,m)}] \notag \\
+& J^\prime_x \sum_{\ell = 1}^{L_x} \sum_{m = 1}^{L_y} \sigma^x_{(6\ell -3,m)} \sigma^x_{(6\ell-4 ,m)}\notag\\
+&J^\prime_y\sum_{\ell = 1}^{L_x-1} \sum_{m = 1}^{L_y}  \sigma^y_{(6\ell,m)} \sigma^y_{(6\ell+1 ,m)}  \notag\\
+&J^\prime_z \sum_{\ell = 1}^{L_x} \sum_{m = 1}^{L_y}  \sigma^z_{(6\ell -1,m)} \sigma^z_{(6\ell-4 ,m)}.
\label{eq:Hamiltonian_trianglehoneycomb_spin}
\end{align}
See Fig.~\ref{Fig:TH} for the geometry and the interactions.
We perform the Jordan-Wigner transformation of Eqs. (\ref{eq:JW_sigmax})-(\ref{eq:JW_sigmaz}),
and decompose the complex fermion $a_{i}$ into two Majorana fermions 
$c_{i}$ and $d_{i}$ in the same way as that in Eqs. (\ref{eq:Majoranaodd}) and (\ref{eq:Majoranaeven}).
Using these transformations for Eq. (\ref{eq:Hamiltonian_trianglehoneycomb_spin}), we obtain
\begin{align}
H =  & i J_x \sum_{\ell = 1}^{L_x} \sum_{m = 1}^{L_y}[ c_{(6\ell -5,m)} c_{(6\ell-4 ,m)} + c_{(6\ell -1,m)} c_{(6\ell ,m)}  ] \notag \\
+& i J_y\sum_{\ell = 1}^{L_x} \sum_{m = 1}^{L_y} [c_{(6\ell -4,m)} c_{(6\ell-3 ,m)} + c_{(6\ell -2,m)} c_{(6\ell-1 ,m)} ]  \notag \\
+ &J_z \sum_{\ell = 1}^{L_x} \sum_{m = 1}^{L_y}  [c_{(6\ell -5,m)} c_{(6\ell-3 ,m)} d_{(6\ell -5,m)} d_{(6\ell-3 ,m)} \notag \\
+&  c_{(6\ell -2,m)} c_{(6\ell ,m)} d_{(6\ell -2,m)} d_{(6\ell ,m)} ] \notag \\
+&  i J^\prime_x \sum_{\ell = 1}^{L_x} \sum_{m = 1}^{L_y} c_{(6\ell -3,m)} c_{(6\ell-4 ,m)} \notag \\
+ & i J^\prime_y\sum_{\ell = 1}^{L_x-1} \sum_{m = 1}^{L_y}  c_{(6\ell,m)} c_{(6\ell+1 ,m)} \notag \\
+ &J^\prime_z \sum_{\ell = 1}^{L_x} \sum_{m = 1}^{L_y}  c_{(6\ell -1,m)} c_{(6\ell-4 ,m)} d_{(6\ell -1,m)} d_{(6\ell -4,m)}.
\label{eq:Hamiltonian_trianglehoneycomb_d}
\end{align}
Again, the pairs $d_i d_j$ in Eq. (\ref{eq:Hamiltonian_trianglehoneycomb_d}) 
commute with the Hamiltonian and thus can be replaced with classical numbers.
We set the link variables as
$d_{(6\ell -2,m)} d_{(6\ell ,m)} = d_{(6\ell -1,m)} d_{(6\ell -4,m)} = +i$,
and $d_{(6\ell -5,m)} d_{(6\ell -3,m)} = -i$, 
which gives us the ground state~\cite{Yao2007}. 
This choice of the link variables leads to the time-reversal symmetry breaking.
In fact, the time-reversal operation changes the sign of the link variables ($\pm i \rightarrow \mp i$),
which gives another state with the same energy. Therefore, the ground state has at least twofold degeneracy
due to the time-reversal symmetry breaking~\cite{Yao2007}.
Then, the Hamiltonian for $c$-Majorana fermions is
\begin{align}
H =  & i J_x \sum_{\ell = 1}^{L_x} \sum_{m = 1}^{L_y}[ c_{(6\ell -5,m)} c_{(6\ell-4 ,m)} + c_{(6\ell -1,m)} c_{(6\ell ,m)}  ]\notag \\
+& i J_y\sum_{\ell = 1}^{L_x} \sum_{m = 1}^{L_y} [c_{(6\ell -4,m)} c_{(6\ell-3 ,m)} + c_{(6\ell -2,m)} c_{(6\ell-1 ,m)} ]  \notag \\
+ & i J_z \sum_{\ell = 1}^{L_x} \sum_{m = 1}^{L_y}  [- c_{(6\ell -5,m)} c_{(6\ell-3 ,m)} + c_{(6\ell -2,m)} c_{(6\ell ,m)} ]\notag \\
+&  i J^\prime_x \sum_{\ell = 1}^{L_x} \sum_{m = 1}^{L_y} c_{(6\ell -3,m)} c_{(6\ell-4 ,m)}\notag \\
+ &  i J^\prime_y\sum_{\ell = 1}^{L_x-1} \sum_{m = 1}^{L_y}  c_{(6\ell,m)} c_{(6\ell+1 ,m)} \notag \\
+ & i  J^\prime_z \sum_{\ell = 1}^{L_x} \sum_{m = 1}^{L_y}  c_{(6\ell -1,m)} c_{(6\ell-4 ,m)}.
\label{eq:Hamiltonian_trianglehoneycomb}
\end{align}
To rewrite Eq. (\ref{eq:Hamiltonian_trianglehoneycomb}) in the BdG form, 
we define three species of complex fermions:
\begin{subequations}
 \begin{equation}
\alpha_{(\ell,m), A} =  \frac{1}{2} [c_{(6 \ell -5 , m)} + i c_{ (6 \ell -4,m) } ], 
\end{equation}
 \begin{equation}
\alpha_{(\ell,m), B} =  \frac{1}{2} [c_{ (6 \ell -3 , m) } + i c_{ (6\ell -2,m) } ], 
\end{equation}
and 
\begin{equation}
\alpha_{(\ell,m), C} =  \frac{1}{2} [c_{(6 \ell -1 , m)} + i c_{(6\ell ,m)} ], 
\end{equation}
\end{subequations}
and their Fourier transformations,
\begin{eqnarray} 
\alpha_{\ell,k_y,s} = \frac{1}{\sqrt{L_y}}\sum_{m=1}^{L_y} \alpha_{(\ell,m), s} e^{ ik_y m}, 
\end{eqnarray}
for $s = A,B,C$. 
Then, the Hamiltonian of Eq. (\ref{eq:Hamiltonian_trianglehoneycomb}) is rewritten as
\begin{align}
H =  \sum_{k_y} \sum_{\ell, \ell^\prime} 
\bm{\Psi}^\dagger_{\ell} (k_y) 
[\hat{h}  (k_y)]_{\ell, \ell^\prime} \bm{\Psi}_{\ell^\prime}(k_y),  \label{eq:ham_trihoney_bdg}
\end{align}
\begin{widetext}
where $\bm{\Psi}_{\ell} (k_y) = (\alpha_{\ell,k_y,A}, \alpha_{\ell,k_y,B},\alpha_{\ell,k_y,C},\alpha^\dagger_{\ell,-k_y,A},\alpha^\dagger_{\ell,-k_y,B},\alpha^\dagger_{\ell,-k_y,C})^{\rm T} $
and 
\begin{eqnarray}
[\hat{h}  (k_y)]_{\ell, \ell^\prime} 
=  \delta_{\ell,\ell^\prime}  M + \delta_{\ell, \ell^\prime-1} \Gamma + 
+ \delta_{\ell, \ell^\prime + 1} \Gamma^\dagger,
\end{eqnarray}
with
\begin{eqnarray}
M = \left(
\begin{array}{cccccc}
J_x & \frac{- J_y - iJ_z}{2}&\frac{J_z^\prime e^{-ik_y}}{2} & 0 & \frac{-J_y - iJ_z}{2} &\frac{J_z^\prime e^{- i k_y}}{2}  \\
 \frac{- J_y +  iJ_z}{2}&J^\prime_x  &  \frac{- J_y + iJ_z}{2}&  \frac{J_y + iJ_z}{2}  & 0 & \frac{-J_y - iJ_z}{2}  \\
 \frac{J_z^\prime e^{ik_y}}{2}&   \frac{- J_y - iJ_z}{2}& J _x&  \frac{-J_z^\prime e^{i k_y}}{2} &  \frac{J_y + iJ_z}{2}  &0 \\
0 &  \frac{J_y - iJ_z}{2}  & \frac{-J_z^\prime e^{- i k_y}}{2} &  -J_x& \frac{ J_y - iJ_z}{2} & \frac{-J_z^\prime e^{-i k_y}}{2} \\
 \frac{-J_y +  iJ_z}{2} &0 &  \frac{J_y - iJ_z}{2} & \frac{J_y + iJ_z}{2} & -J^\prime_x &\frac{J_y + iJ_z}{2}  \\
\frac{J_z^\prime e^{ i k_y}}{2}&  \frac{-J_y + iJ_z}{2} & 0& \frac{-J_z^\prime e^{i k_y}}{2} & \frac{ J_y - iJ_z}{2} &- J_x\\
\end{array}
\right),
\end{eqnarray}
and 
\begin{eqnarray}
\Gamma = \left(
\begin{array}{cccccc}
0 & 0 &0 & 0 & 0 &0 \\
0 & 0 &0 & 0 & 0 & 0\\
 \frac{-J_y^\prime}{2} & 0 &0 & \frac{-J_y^\prime}{2} & 0 &0 \\
0 & 0 &0 & 0 & 0 &  0\\
0 & 0 &0 & 0 & 0 &0 \\
\frac{J_y^\prime}{2} & 0 &0 & \frac{J_y^\prime}{2} & 0 & 0\\
\end{array}
\right).
\end{eqnarray}
\end{widetext}
The topological class of this Hamiltonian is also D class, and hence the model exhibits a chiral edge mode.

\section{Dispersion relation of edge mode \label{sec:chiraledge_transfer}}
\begin{figure}[b]
\begin{center}
\includegraphics[width= 0.95\linewidth]{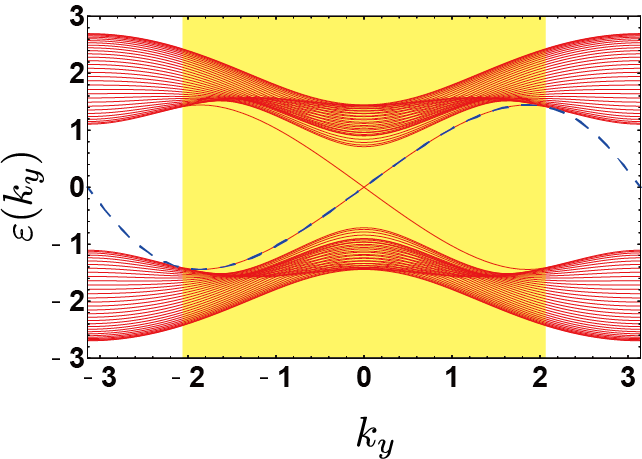}
\caption{The dispersion relation of the BdG Hamiltonian in of 
Eq. (\ref{eq:BDG}) for $(J_x,J_y,J_z,J^\prime) = (1,0.8,0.9,0.3)$.
The red solid lines denote the energy spectrum obtained by the numerical 
diagonalization for the Hamiltonian, and the blue dashed line denotes 
the dispersion of the edge mode which is analytically obtained in the text. 
Only the yellow area allows the decay edge solution which satisfies 
the decay condition $|\lambda(k_y)|<1$. }
\label{Fig2}
\end{center}
\end{figure}
\subsection{Kitaev honeycomb model}
In this section, we derive the dispersion relation of the chiral edge mode 
at the left edge for the Kitaev model. To obtain the exact solution of the 
chiral edge mode, we employ a transfer matrix method. 
As we mentioned in Sec.~\ref{sec:intro}, our method is applicable to generic 
models described by $4\times4$ transfer matrices.

The left edge mode $\gamma^{\mathrm{L}}(k_y)$ can be expanded in terms of $\bm{\Psi}_\ell (k_y)$ as 
\begin{eqnarray}
\gamma^{\mathrm{L}}(k_y) =
\sum_{\ell = 1}^{L_x} \bm{\varphi}^{\mathrm{T}}_{\ell} (k_y) \cdot \bm{\Psi}_\ell (k_y),  \label{eq:leftedge}
\end{eqnarray} 
with 
\begin{eqnarray}
\bm{\varphi}_{\ell} (k_y) =
\left(
\begin{array}{c}
u_{\ell, k_y} \\
v_{\ell, k_y} \\
\end{array}
\right).
\end{eqnarray}
Note that $\gamma^{\mathrm{L}}(k_y)$ satisfies $\left[ \gamma^{\mathrm{L}}(k_y), H \right] = \varepsilon^{\mathrm{L}} (k_y) \gamma^{\mathrm{L}}(k_y)$. 
This leads to the eigenvalue equation 
for $\bm{\varphi}_{\ell} (k_y)$,
which is give by
\begin{eqnarray}
&& \hat{A}(k_y) \bm{\varphi}_{\ell+1} (k_y) + \hat{A}^\dagger (k_y)\bm{\varphi}_{\ell-1} (k_y) + \hat{B}(k_y) \bm{\varphi}_{\ell} (k_y) \nonumber \\
&=& \varepsilon^{\mathrm{L}} (k_y)  \bm{\varphi}_{\ell} (k_y), \label{eq:eigen}
\end{eqnarray}
where $\hat{A}(k_y)$ and $\hat{B}(k_y)$ are $2 \times 2 $ matrices given as 
\begin{eqnarray}
\hat{A}(k_y)=\left(
\begin{array}{cc}
-\frac{J_y}{2} & \frac{J_y}{2} + iJ^\prime (e^{i k_y} + 1) \\
-\frac{J_y}{2} + iJ^\prime (e^{i k_y} + 1) & \frac{J_y}{2} \\
\end{array}
\right), \nonumber \\
\end{eqnarray}
and
\begin{eqnarray}
\hat{B}(k_y)=\left(
\begin{array}{cc}
J_x -J_z \cos k_y & -i J_z \sin k_y  - 2J^\prime \sin k_y\\
 i J_z \sin k_y  - 2 J^\prime \sin k_y& -J_x  + J_z \cos k_y  \\
 \end{array}
\right).\nonumber \\
\end{eqnarray}
Equation (\ref{eq:eigen}) can be rewritten as
\begin{eqnarray}
\left(
\begin{array}{c}
\bm{\varphi}_{\ell + 1} \\
\bm{\varphi}_{\ell} \\
\end{array}
\right)
 = \hat{T}(k_y) \left(
\begin{array}{c}
\bm{\varphi}_{\ell } \\
\bm{\varphi}_{\ell -1 } \\
\end{array}
\right)  \label{eq:TM_eq}
\end{eqnarray}
in terms of the $4 \times 4$ transfer matrix,
\begin{eqnarray}
\hat{T} (k_y) := \left(
\begin{array}{cc}
\hat{A}^{-1}(k_y) [\varepsilon^{\mathrm{L}} (k_y)-\hat{B}(k_y)]  & - \hat{A}^{-1}(k_y) \hat{A}^{\dagger}(k_y) \\
\hat{I}_2 & 0\\
\end{array}
\right),\nonumber \\ \label{eq:TM_def}
\end{eqnarray}
where $\hat{I}_2$ stands for the $2 \times 2$ identity matrix. 

From Eqs. (\ref{eq:TM_eq}) and (\ref{eq:TM_def}), 
one finds that
the energy eigenvalue $ \varepsilon^{\mathrm{L}} (k_y)$ has to be determined 
such that $\hat{A}^{-1}(k_y) [\varepsilon^{\mathrm{L}} (k_y)-\hat{B}(k_y)]$
and $- \hat{A}^{-1}(k_y)\hat{A}^{\dagger}(k_y)$ have simultaneous eigenstates;
see Appendix \ref{sec:edgemode_method} for details.
After some calculations, we obtain the dispersion relation as
\begin{eqnarray}
\varepsilon^{\mathrm{L}} (k_y) = 
\frac{2J^\prime \left(J_x +J_y +J_z \right) \sin k_y}{ \sqrt{J_y^2 + 8J^{\prime 2} \left(1+\cos k_y \right)} }. 
\label{eq:edge_disp}
\end{eqnarray}
Notice that, in order that the wave function $\bm{\varphi}_{\ell}$ is localized at the left edge, 
the eigenvalues of $\hat{T}(k_y)$, $\lambda(k_y)$, 
has to satisfy the condition $|\lambda(k_y)| < 1$. 

In Fig. \ref{Fig2}, we plot the energy eigenvalues of the Hamiltonian of Eq. (\ref{eq:BDG}) (red solid lines),
together with the edge-mode dispersion given in Eq. (\ref{eq:edge_disp}) (a blue dashed line),
showing quite good agreement with each other. 
It can also be seen in Fig. \ref{Fig2} that the chiral edge mode has a linear dispersion near $k_y=0$.
The velocity, which is defined as
\begin{eqnarray}
v^{\mathrm{L}}_{\mathrm{g}} (k_y)  = \frac{\partial \varepsilon^{\mathrm{L}} (k_y)}{\partial k_y},
\nonumber \\
 \label{eq:groupvelo}
\end{eqnarray}
is approximated in the vicinity of $k_y=0$ as
\begin{eqnarray}
v^{\mathrm{L}}_{\mathrm{g}} (k_y \sim 0)  \sim  \frac{2J^\prime \left(J_x +J_y +J_z \right) }{ \sqrt{J_y^2 + 16J^{\prime 2} } }.
\nonumber \\
 \label{eq:groupvelo2}
\end{eqnarray}

\subsection{Kitaev triangle-honeycomb model}
\begin{figure}[t]
\begin{center}
\includegraphics[width=\linewidth]{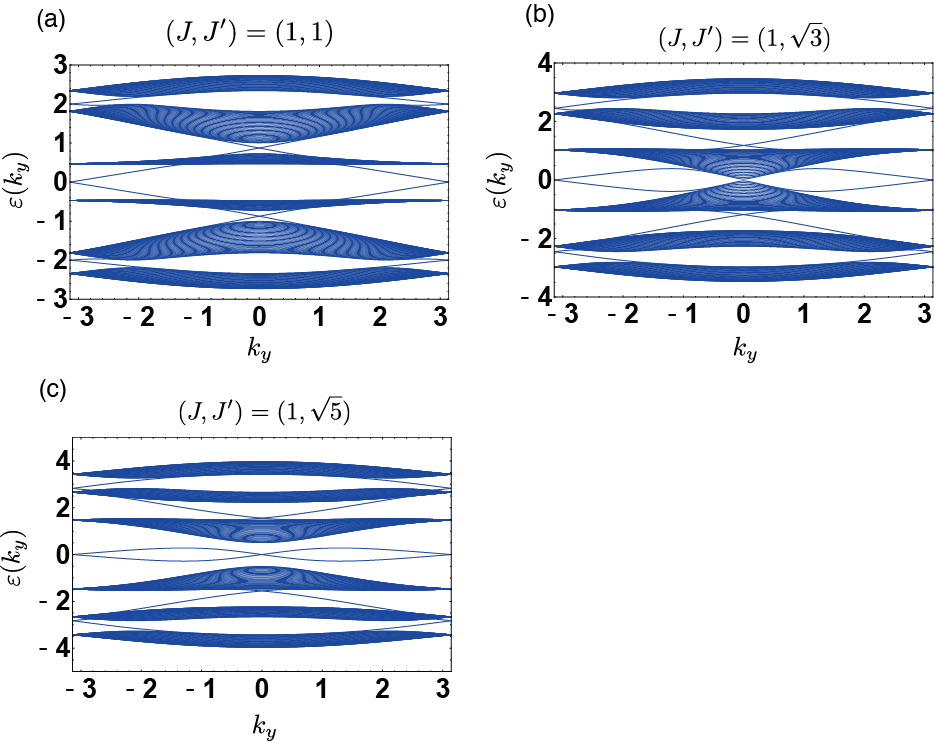}
\caption{
Dispersion relations for the Kitaev triangle-honeycomb model
for (a) $(J,J^\prime ) = (1,1)$, (b) $(J,J^\prime ) = (1,\sqrt{3})$, and (c) $(J,J^\prime ) = (1,\sqrt{5} )$.
}
\label{Fig7}
\end{center}
\end{figure}
Next, let us treat the Kitaev triangle-honeycomb model.
Following Ref.~\onlinecite{Yao2007},
we restrict the parameters to $J_x = J_y =J_z = J$, and 
$J_x^\prime = J_y^\prime =J_z^\prime = J^\prime$. 
It has been pointed out that the ground state is topologically nontrivial (i.e. having a finite Chern number) for $|J^\prime| < \sqrt{3}|J|$,
and is trivial (i.e., having zero Chern number) for $|J^\prime| > \sqrt{3}|J|$; $|J^\prime| = \sqrt{3}|J|$ is a critical point at which the bulk band gap closed. 

Although the transfer matrix formalism shown in the previous subsection is also applicable 
to the present model, it is fairly difficult to obtain the exact solution, since the size of matrices is large. 
We therefore employ a different method to obtain the edge modes, which is similar to one used in the prior work~\cite{Mizoguchi2019};
all the details of that method are described in Appendix~\ref{sec:trihoney}.

Here we look at the numerical results:
In Figs.~\ref{Fig7}(a)-(c), we plot the dispersion relations for the BdG equation of Eq. (\ref{eq:ham_trihoney_bdg}), 
obtained by the numerical diagonalization of the Hamiltonian.
For $(J,J^\prime) = (1,1)$, there exist chiral edge modes, 
originating from the nonzero Chern number of the bulk band. 
Interestingly, even for a nontopological phase with vanishing Chern number [$(J,J^\prime) = (1,\sqrt{5})$], 
there still exist the edge states, but they are not chiral,
because the velocity around $k_y = \pm \pi$ takes the opposite sign to that around $k_y =0$.
This leads to the localization of the edge modes in the presence of the impurities due to the back scattering, 
i.e., the edge modes in the nontopological phase are not stable against disorders unlike the chiral edge modes~\cite{Elbau2002} in the topological phase.

\section{Propagation of magnetization at the edge \label{sec:edgemag_propagate}}
In this section, we calculate the time evolution of the edge magnetization. 
To do this, we combine the technique to calculate the edge magnetization in the previous work~\cite{Mizoguchi2019}
with the result of the eigenvalues and eigenvectors of the left edge mode derived in the previous section.

As a initial state of dynamics, we consider the situation where the chiral edge mode are excited 
with a certain weight function $\rho_{k_y}$ which is a function of the vertical momentum $k_y$. 
Physically, this situation can be achieved by applying an external perturbation, e.g., 
shining laser pulse to the sample magnet whose energy is lower than the bulk energy gap of the magnet. 
We assume that such an external perturbation is weak enough not to excite the flux sector. 

\subsection{Formulation}
\subsubsection{Warm-up: A spinless superconductor}
Consider first the case that 
the link variable, $d_{(2\ell,m)}d_{(2\ell-1,m+1)}$,
is equal to $+i$ for all $(\ell,m)$.  
Then, the Hamiltonian (\ref{eq:ham_trihoney_bdg}) can be also interpreted as the Hamiltonian of a spinless superconductor.
Therefore, before proceeding to the calculation of the edge magnetization
in the Kitaev-type models, we compute the propagation of a Majorana edge charge excitation
in the superconductor. The calculation is much simpler than those in the Kitaev-type models.

To begin with, we recall 
\begin{eqnarray}
\alpha_{\ell,k_y}= \frac{1}{\sqrt{L_y}} 
\sum_{m=1}^{L_y}e^{ik_y m} \frac{c_{(2\ell-1,m)} + ic_{(2\ell,m)}}{2}. \label{eq:alpha_c}
\end{eqnarray}
Substituting (\ref{eq:alpha_c}) into Eq. (\ref{eq:leftedge}),
we obtain
\begin{eqnarray}
\gamma^{\mathrm{L}} (k_y)
= \frac{1}{\sqrt{L_y}} \sum_{\ell=1}^{L_x}\sum_{m=1}^{L_y}
e^{ik_ym}
\left[\frac{\zeta_{\ell,k_y}}{2}c_{(2\ell-1,m)}+i\frac{\xi_{\ell,k_y}}{2}c_{(2\ell,m)}\right], \nonumber \\ \label{eq:gamma_c}
\end{eqnarray}
with $\zeta_{\ell,k_y}=u_{\ell,k_y}+v_{\ell,k_y}$ and $\xi_{\ell,k_y}=u_{\ell,k_y}-v_{\ell,k_y}$. 

The time evolution of $\gamma^{\mathrm{L}} (k_y)$ is given by
\begin{eqnarray}
\gamma^{\mathrm{L}} (k_y, t) := e^{-iH t} \gamma^{\mathrm{L}} (k_y)e^{iH t}
= e^{i \varepsilon^{\mathrm{L}}(k_y) t} \gamma^{\mathrm{L}} (k_y). 
\end{eqnarray}
Using $\gamma^{\mathrm{L}} (k_y, t)$, 
let us consider the wave packet created by the aforementioned external perturbation, 
\begin{eqnarray}
\eta[\rho](t):=\frac{1}{\sqrt{L_y}}\sum_{k_y} \rho_{k_y}\left[\gamma^{\mathrm{L}}(k_y,t)+\gamma^{\mathrm{L} \dagger}(-k_y,t) \right],\label{eq:etat}
\end{eqnarray}
with the weight function $\rho_{k_y}$ which satisfies $\rho_{-k_y}^\ast=\rho_{k_y}$. 
One can easily show $\eta[\rho]^\dagger(t)=\eta[\rho](t)$. 
Noting that the dispersion relation given by Eq. (\ref{eq:edge_disp})
satisfies $\varepsilon^{\mathrm{L}}(k_y) = - \varepsilon^{\mathrm{L}}(-k_y)$, one obtains 
\begin{eqnarray}
\eta[\rho](t)=\sum_{n=1}^{2L_x}\sum_{m=1}^{L_y}\frac{1}{L_y}
\sum_{k_y} \rho_{k_y}{\cal A}_{n,k_y}e^{ik_y m}e^{i\varepsilon^{\rm L} (k_y)t}c_{(n,m)}, \nonumber \\
\end{eqnarray}
by substituting Eq. (\ref{eq:gamma_c}) into Eq. (\ref{eq:etat}). 
Here we have introduced 
\begin{equation}
{\cal A}_{n,k_y}:=
\left\{
\begin{array}{cc}
(\zeta_{\ell,k_y}+\zeta_{\ell,-k_y}^\ast)/2, & \mathrm{for} \hspace{1mm} n=2\ell-1; \\
i(\xi_{\ell,k_y}-\xi_{\ell,-k_y}^\ast)/2, & \mathrm{for} \hspace{1mm} n=2\ell. 
\end{array} 
\right.
\end{equation}
For ${\cal A}_{n,k_y}$, one can show the relation, 
\begin{equation}
{\cal A}_{n,-k_y}^\ast={\cal A}_{n,k_y}.
\label{calArelation}
\end{equation}

We further define
\begin{equation}
\psi_{n,m}(t):=\frac{1}{L_y}\sum_{k_y}
{\tilde{\cal A}}_{n,k_y}e^{ik_y m}e^{i\varepsilon^{\rm L} (k_y)t}.
\label{psit}
\end{equation}
with ${\tilde{\cal A}}_{n,k_y}=\rho_{k_y}{\cal A}_{n,k_y}$. 
Then, using the above relation (\ref{calArelation}), as well as
$\rho_{-k_y}^\ast=\rho_{k_y}$ and $\varepsilon^{\rm L} (-k_y)=-\varepsilon^{\rm L} (k_y)$, 
one has $\psi_{n,m}^\ast(t)=\psi_{n,m}(t)$, 
i.e., $\psi_{n,m}(t)$ is a real function. 
Clearly, one has 
\begin{equation}
\eta[\rho](t)=\sum_{n=1}^{2L_x}\sum_{m=1}^{L_y}\psi_{n,m}(t)c_{(n,m)}.
\label{etarhotc}
\end{equation}
The wave function $\psi_{n,m}(t)$ is localized at $n$. 
In the limit $L_y\rightarrow \infty$, the wave function is written as
\begin{equation}
\psi_{n,m}(t)=\int \frac{d k_y}{2\pi} \; {\tilde{\cal A}}_{n,k_y}e^{ik_y m}e^{i\varepsilon^{\rm L} (k_y)t}. \label{eq:psi}
\end{equation}
Therefore, if the Fourier component ${\tilde{\cal A}}_{n,k_y}$ 
is a smooth function with respect to $k_y$, then $\psi_{n,m}(t)$ is localized at $m$, too.  
Note that 
\begin{eqnarray}
m\psi_{n,m}(t)&=&\int \frac{d k_y}{2\pi} \; {\tilde{\cal A}}_{n,k_y}m e^{ik_y m}e^{i\varepsilon^{\rm L} (k_y)t} \nonumber \\
&=&\int \frac{dk_y}{2\pi} \; {\tilde{\cal A}}_{n,k_y}\left(-i\frac{\partial}{\partial k_y} e^{ik_y m}\right)
e^{i\varepsilon^{\rm L} (k_y)t} \nonumber  \\
&=&\int \frac{dk_y}{2\pi} \;e^{ik_y m} e^{i\varepsilon^{\rm L} (k_y)t}\left(i\frac{\partial {\tilde{\cal A}}_{n,k_y} }{\partial k_y}
-{\tilde{\cal A}}_{n,k_y}\frac{\partial \varepsilon^{\rm L} (k_y)}{\partial k_y} t\right). \nonumber  \\
\end{eqnarray}
Therefore, the expectation value of the position $m$ is given by 
\begin{eqnarray}
&& \sum_{m=1}^{L_y}m|\psi_{n,m}(t)|^2 \nonumber \\
&=&\int \frac{d k_y}{2\pi}
\left[{\tilde{\cal A}}_{n,k_y}^\ast i\frac{\partial}{\partial k_y}{\tilde{\cal A}}_{n,k_y}
-\left|{\tilde{\cal A}}_{n,k_y}\right|^2 v^{\mathrm{L}}_{\mathrm{g}}(k_y)t\right], \nonumber \\
\end{eqnarray}
where $v^{\mathrm{L}}_{\rm g}(k_y)$ is given in Eq. (\ref{eq:groupvelo}). 
This implies that the wave packet propagates with the group velocity 
\begin{eqnarray}
{\cal V}_{\rm g}=\frac{\int dk_y\;\left|{\tilde{\cal A}}_{1,k_y}\right|^2v^{\mathrm{L}}_{\rm g}(k_y) }
{\int dk_y\;\left|{\tilde{\cal A}}_{1,k_y}\right|^2} \label{eq:GV}
\end{eqnarray}
at the left edge of the sample. 
Since $v^{\mathrm{L}}_{\rm g}(k_y)$ in Eq. (\ref{eq:groupvelo}) 
is always positive (or negative) irrespective of the wave number $k_y$,
so is the group velocity ${\cal V}_{\rm g}$. 
This implies that an oriented propagation is realized for the chiral edge mode. 

In order to obtain the oriented propagation of the Majorana edge charge excitation, we consider a trial state,
\begin{eqnarray}
|\Psi(t) \rangle = \mathcal{N} \{1 + \eta[\rho](t) \} |0 \rangle,  \label{eq:trialwf_1}
\end{eqnarray}
where $\mathcal{N}$ is the normalization factor.
Note that the wave function of Eq. (\ref{eq:trialwf_1}) 
is chosen such that the states with different fermion parities are mixed. 
This is the standard technique to obtain the finite expectation value of a single Majorana fermion~\cite{Mizoguchi2019,Willans2010,Willans2011}, as shown below. 
The expectation value of the Majorana fermion $c_{(1,m)}$ can be calculated as
\begin{eqnarray}
 && \langle \Psi(t)  | c_{(1,m)} | \Psi(t) \rangle \nonumber \\
&=& |\mathcal{N}|^2 \langle 0 |\{1+\eta^\dagger[\rho](t)\}c_{(1,m)}\{1+\eta[\rho](t)\}|0\rangle \nonumber \\
&=& |{\cal N}|^2 \langle 0|\{\eta[\rho](t)c_{(1,m)}+c_{(1,m)}\eta[\rho](t)\}|0\rangle  \nonumber \\
&=& |{\cal N}|^2 \sum_{n,m^\prime} \psi_{n,m^\prime}(t) \langle 0 |\{c_{(n,m^\prime)}, c_{(1,m)} \} | 0 \rangle  \nonumber \\
&=& 2 |{\cal N}|^2 \psi_{1,m}(t), 
\end{eqnarray}
where we have used $\eta^\dagger[\rho](t)=\eta[\rho](t)$. Thus, the profile of the excess charge of the Majorana edge 
fermion is expressed in terms of the amplitude $\psi_{1,m}(t)$ of the wave function at the edge site (1,m), 
and the charge moves in one of the two directions of the edge with the group velocity ${\cal V}_{\rm g}$ of (\ref{eq:GV}). 

\subsubsection{Kitaev-type models \label{sec:generic}}
Now, to realize the propagation of the edge magnetization, 
we need to consider  
a slightly generic trial state. 
As we have shown in the previous work~\cite{Mizoguchi2019}, 
the operator $e^{i\pi\hat{\theta}_{(1,m)}}$ plays a crucial role 
for the calculation of the edge magnetization,
since the spin operators $\bm{\sigma}_{(1,m)}$ inevitably contain it. 
Since the operator $e^{i\pi\hat{\theta}_{(1,m)}}$ commutes with the Hamiltonian $H$, 
the state $e^{i\pi\hat{\theta}_{(1,m)}}|0\rangle$ is still a ground state for the fermion vacuum state $|0\rangle$. 
But the operation changes the link variable, $d_{(2\ell,m)}d_{(2\ell-1,m+1)}$, for some $m$.  
This causes the change of the signs of the hopping amplitudes of $J_z$ bonds for $c$-Majorana fermions. 
Of course, this change of the signs can be removed by using the gauge transformation  
$c_{(2\ell-1,m)}\rightarrow \pm c_{(2\ell-1,m)}$. 

In this case, instead of the trial state of (\ref{eq:trialwf_1}), we consider a trial excited state, 
\begin{eqnarray}
|\Psi(t)\rangle=\frac{1}{\sqrt{2}}
\left\{1+W \eta[\rho](t)\right\}|0\rangle, 
\end{eqnarray}
where the operator $W$ is defined by 
\begin{eqnarray}
W:=w(1)+\sum_{m=2}^{L_y} w(m)e^{i\pi\hat{\theta}_{(1,m)}} \label{eq:weightw}
\end{eqnarray}
with a positive weight function $w(m)$ which satisfies the normalization 
condition,
\begin{eqnarray}
\sum_{m=1}^{L_y} |w(m)|^2=1. \label{eq:w_normalize}
\end{eqnarray}
We take the wave function $\psi_{n,m}(t)$ of (\ref{eq:psi}) 
and the weight function $w(m)$ so that the overlap between 
them at the initial time $t=0$ is large. 
As we will show below, it is essential to set $w(m)$ as a slowly-decaying function.
For instance, if $w(m)$ is set to be constant, then Eq. (\ref{eq:w_normalize}) requires that $w(m) = \frac{1}{\sqrt{L_y}}$,
which leads to the vanishing of the edge magnetization for $L_y \rightarrow \infty $ [see Eq. (\ref{eq:edgemag_finresult})].

Using the parity conservation for the fermions, one can check that $|\Psi(t)\rangle$ is normalized as  
\begin{widetext}
\begin{eqnarray}
\Vert\Psi(t)\Vert^2&=&\frac{1}{2} 
+ \frac{1}{2}\langle 0| \eta[\rho] (t) \left[w(1)+\sum_{m=2}^{L_y} w(m)e^{i\pi\hat{\theta}_{(1,m)}}\right]^2 \eta[\rho] (t)|0\rangle \nonumber \\
&=&\frac{1}{2} + \frac{1}{2}\sum_{m=1}^{L_y}|w(m)|^2=1.
\end{eqnarray}
\end{widetext}
Here, we also have used that the operators $e^{i\pi\hat{\theta}_{(1,m)}}$ 
change the link variables and that two states with different link variables are orthogonal with 
each other, and we have taken the wave function of (\ref{eq:psi}) to be normalized to 1. 

Let us compute the expectation value of the $y$-component spin at the site $(1,m)$ which 
is given by $\langle\Psi(t)|\sigma_{(1,m)}^y|\Psi(t)\rangle$. 
We note that the other two components ($x$ and $z$) have vanishing expectation values for $\ket{\Psi(t)}$ 
because they contain a single $d$-Majorana fermion operator except for the string operator.
To obtain finite values of them, one may need to set the trial wave function 
to be a linear combination of wave functions with different fermion parities of the $d$-Majorana fermions.
However, it seems to be fairly difficult not only to find such a favorable wave function, but also to compute the expectation values.
The $y$ component of the spin is written  
\begin{eqnarray} 
\sigma_{(1,m)}^y=c_{(1,m)}e^{i\pi\hat{\theta}_{(1,m)}} \label{eq:sigmay}
\end{eqnarray}
in terms of the Majorana fermion $c_{(1,m)}$. In the same way as the above calculation, 
by using this expression, the expression (\ref{etarhotc}) of $\eta[\rho](t)$ and $\{\eta[\rho](t)\}^\dagger=\eta[\rho](t)$, 
we obtain 
\begin{widetext}
\begin{eqnarray}
\mu(m,t):= \langle\Psi(t)|\sigma_{(1,m)}^y|\Psi(t)\rangle
&=&
\frac{1}{2}\langle 0|   \left\{ \eta[\rho](t) \left[w(1)+\sum_{m'=2}^{\ell_y} w(m')e^{i\pi\hat{\theta}_{(1,m')}}\right]
c_{(1,m)}e^{i\pi\hat{\theta}_{(1,m)}} \right\}  \nonumber \\
&+&\left\{ c_{(1,m)}e^{i\pi\hat{\theta}_{(1,m)}}
\left[w(1)+\sum_{m'=2}^{\ell_y} w(m')e^{i\pi\hat{\theta}_{(1,m')}}\right] \eta[\rho](t) \right\}|0\rangle \nonumber \\
&=&\frac{w(m)}{2}\langle 0|\eta[\rho](t)c_{(1,m)} + \left\{c_{(1,m)}\eta[\rho](t) \right\}|0\rangle \nonumber \\
&=&w(m)\psi_{1,m}(t). \label{eq:edgemag_finresult}
\end{eqnarray}
\end{widetext}
We emphasize that the string operator $e^{i\pi\hat{\theta}_{(1,m)}}$ in $W$ is essential to obtain the finite magnetization, 
since the string operator $e^{i\pi\hat{\theta}_{(1,m)}} $ changes the eigenvalues of 
the link variables, $d_{(2\ell,m)}d_{(2\ell-1,m+1)}$, of the fermion vacuum $|0\rangle$ as mentioned above, and yields the vanishing of the magnetization.
In other words, the finite expectation value is obtained by canceling the string operator in $\sigma_{(1,m)}^y$ with those in $W$.
Thus, the profile of the magnetization is expressed in terms of the weight function $w(m)$ and 
the amplitude $\psi_{1,m}(t)$ of the wave function  
at the edge site $(1,m)$. The amplitude slowly decays by the weight $w(m)$ with distance.
Actually, in the limit $L_y \rightarrow\infty$, the magnetization is vanishing 
in the region where $w(m) \rightarrow 0$ [see Eq.~(\ref{eq:weightw2}) as an example].
This implies that it is impossible to suppress the fluctuation of the link variables 
through the whole region in this approach. 
Although the local magnetization decays and finally vanishes, 
it exhibits an oriented propagation along the left edge with the group velocity ${\cal V}_{\rm g}$. 

\begin{figure}[t]
\begin{center}
\includegraphics[width= 0.98\linewidth]{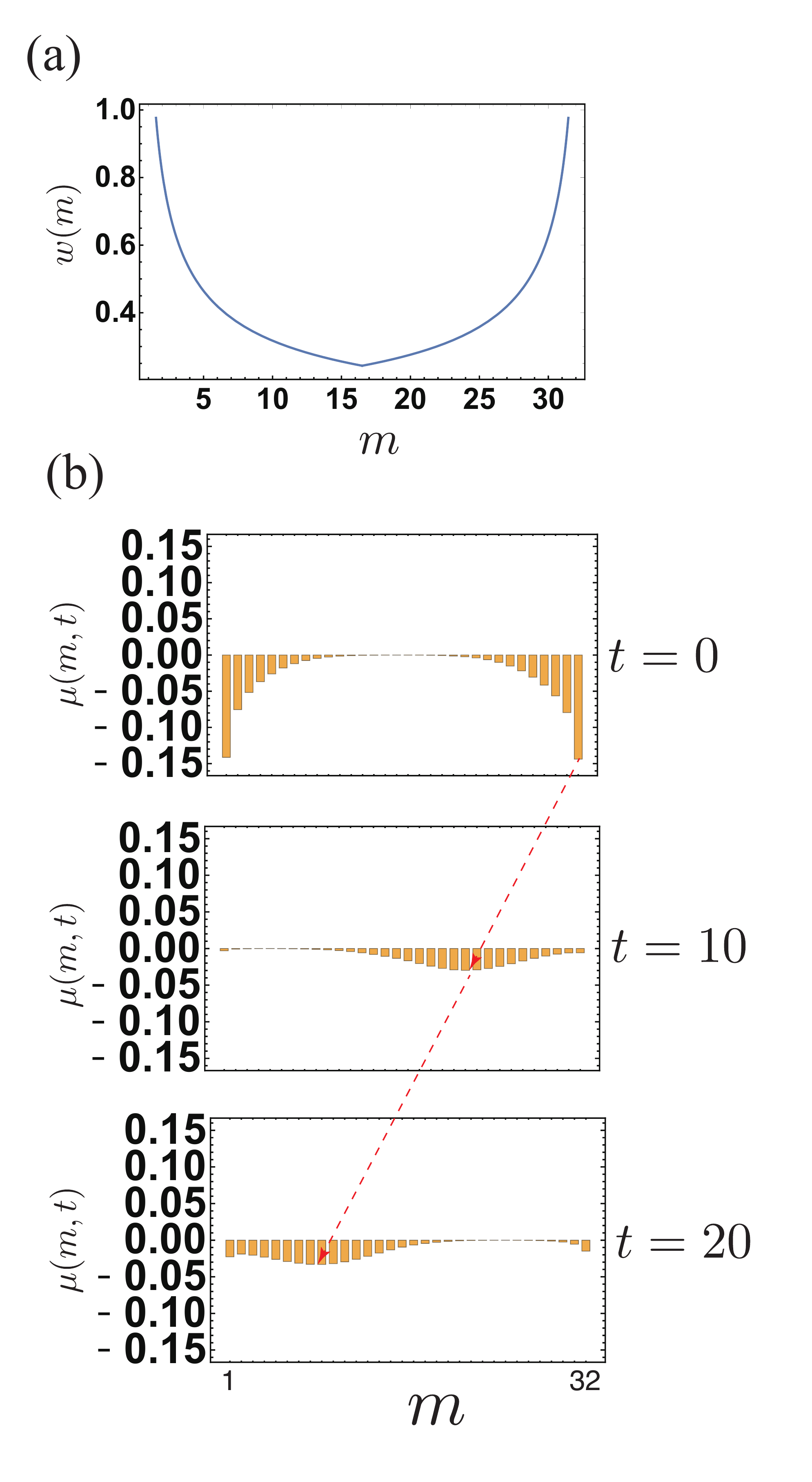}
\caption{
(a) The weight $w(m)$ in Eq. (\ref{eq:weightw2}).
(b) The edge magnetization $\mu(m,t)$ as a function of $m$ for 
$t=0$ (top), $t=10$ (middle), and $t=20$ (bottom)
with $k_0 = 0$. 
The dip of the magnetization propagates from the top row $(m=L_y=32)$ to the bottom row $(m=1)$.
Red arrows are for the guide to the eyes.}
\label{Fig3}
\end{center}
\end{figure}

\subsection{Numerical demonstration}
\subsubsection{Kitaev honeycomb model with three-spin interaction}
Following the formulation shown in the previous subsection, 
we calculate the time evolution of the edge magnetization numerically. 
We first consider the Kitaev honeycomb model with three-spin interaction.
We choose the Gaussian distribution, 
\begin{eqnarray} 
\rho_{k_y} = \frac{1}{\sqrt{\pi k_0}} e^{- \left( \frac{k_y}{k_0} \right)^2 }, \label{eq:weight}
\end{eqnarray}
for the weight function $\rho_{k_y}$ in (\ref{eq:etat}), and  
\begin{eqnarray} 
w(m) = \begin{cases}
\left|m-\frac{1}{2}\right|^{-1/2-\delta} &\mathrm{for}\hspace{1mm}m = 1,2, \cdots, \frac{L_y}{2}, \\
\left|L_y-m+\frac{1}{2}\right|^{-1/2-\delta}  &\mathrm{for}\hspace{1mm}m = \frac{L_y}{2} +1,\cdots, L_y, \\
\end{cases} 
\nonumber \\
\label{eq:weightw2}
\end{eqnarray}
for the weight function $w(m)$ in (\ref{eq:weightw}) [Fig.~\ref{Fig3}(a)].

We plot $\mu(m,t)$ for $(J_x,J_y,J_z, J^\prime) = (1,0.8,0.9,0.3)$ and $L_y=32$, 
with $k_0 = 0.1 \pi$ in Fig.~\ref{Fig3}(b). 
We clearly see that the dip of the magnetization propagates 
from the top row $(m=L_y)$ to the bottom row $(m=1)$ in both cases, 
indicating the oriented propagation of the magnetization due to the chiral edge mode. 
The velocity of the propagating magnetization is estimated from the figure as $1.08$, which is compatible with the 
group velocity given in Eq. (\ref{eq:groupvelo2}), that is, $v_{\mathrm{g}}^{\rm L}(0) \sim 1.12$.  
We also see that the magnetization decays as the center of the wave packet approaches to $m = L_y/2$. 

\subsubsection{Kitaev triangle-honeycomb model}
\begin{figure}[b]
\begin{center}
\includegraphics[width= 0.95\linewidth]{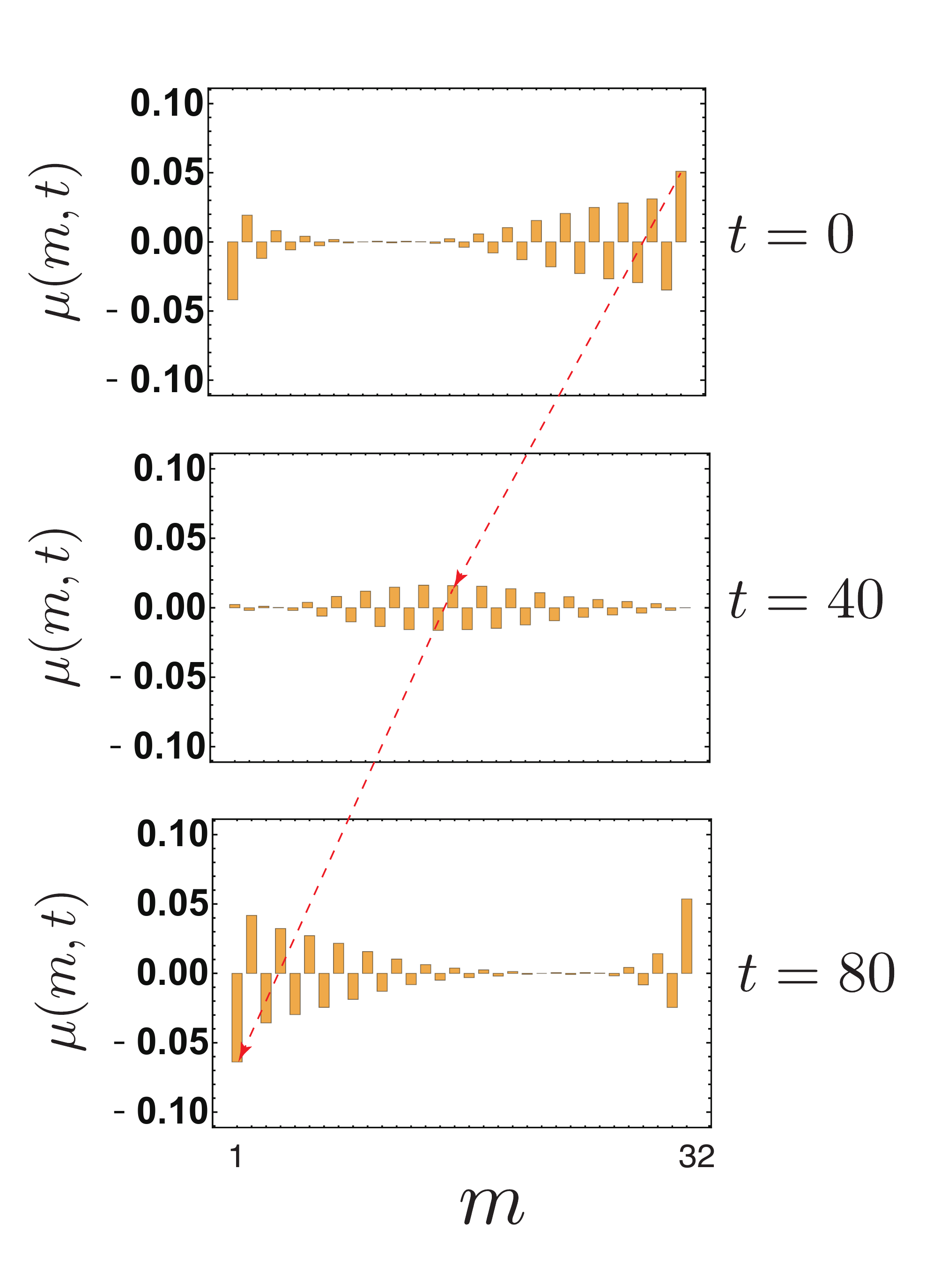}
\caption{
The edge magnetization $\mu(m,t)$ for the Kitaev triangle-honeycomb model as a function of $m$ and for 
$t=0$ (top), $t=40$ (middle), and $t=80$ (bottom). 
Red arrows are a guide for eyes. 
 }
\label{Fig6}
\end{center}
\end{figure}
Next, we consider the Kitaev triangle-honeycomb model.
Practically, we need to replace $\zeta_{1,k_y}$ with $\zeta_{1,k_y,A}$ when applying the formulation to the Kitaev triangle-honeycomb model.
As for the initial state of dynamics, we consider the case where the left edge mode crossing the zero energy is excited.
To choose such a weight function, we need to bear in mind that the edge mode 
in this model exists around $k_y = \pm \pi$ in the momentum space, 
in contrast to the Kitaev honeycomb model with the three-spin interaction, 
where the edge mode exists around $k_y =0$.  
We therefore choose the weight function $\rho_{k_y}$ as 
\begin{eqnarray} 
\rho_{k_y} = \frac{1}{2 \sqrt{\pi k_0}} \left[ e^{- \left( \frac{k_y- \pi }{k_0} \right)^2 } + e^{- \left( \frac{k_y +  \pi }{k_0} \right)^2 } \right], \label{eq:weight2}
\end{eqnarray}
which has a large amplitude at $k_y = \pm \pi$.

In Fig.~\ref{Fig6}, we plot the results for $J=1$, $J^\prime = 1$, $L_y=32$, and $k_0 = 0.1 \pi$.
We see that the magnetization in this system shows the staggered structure in the real space, 
originating from the fact that the weight function has a large amplitude at $k_y = \pm \pi$, in contrast to the Kitaev honeycomb model with the three-spin interaction.
Nevertheless, we see that the magnetization propagates from the top to the bottom, 
due to the existence of the chiral edge spinons. 
Again, we see that the magnetization decays as approaching to $m = L_y/2$,
while the amplitude is recovered near $m \sim 1$ ($t=80$) since $w(m) $ takes large values in that region. 

\section{Edge dynamical susceptibility \label{sec:edge_DS}}
In this section, we calculate the edge structure factor,
\begin{eqnarray}
\chi_{\rm edge} (\omega)=
\frac{1}{L_y} \sum_{m,m^\prime} \int_{-\infty}^{\infty} dt e^{- i \omega t} \langle 0 | \sigma^y_{(1,m)} (t) \sigma^y_{(1,m^\prime)} | 0\rangle,
\nonumber \\ \label{eq:edgeSSF}
\end{eqnarray}
where $|0 \rangle $ stands for the ground state, i.e., 
the state in which all the eigenstates with negative eigenvalues are occupied. 
We consider the situation that the frequency $\omega$ is contained in the region 
of the bulk spectral gap so that only the chiral edge mode is detected. 
Note that we only consider a pair of $\sigma^y$s for the expectation value in Eq. (\ref{eq:edgeSSF}),
since the action of the other two components, $\sigma^x$ and $\sigma^z$, causes the change of the flux sector~\cite{Mizoguchi2019},
which makes the calculation fairly complicated.
As we will explain below, the chiral nature of the edge mode is reflected to this quantity. 

Substituting Eq. (\ref{eq:sigmay}) into the expression (\ref{eq:edgeSSF}) of the susceptibility, 
we have \begin{eqnarray}
&\chi_{\rm edge} (\omega)= \frac{1}{L_y} \sum_{m,m^\prime} \nonumber \\
\times& \int_{-\infty}^{\infty} dt e^{- i \omega t} \langle 0 |  e^{-i H t} c_{(1,m)} e^{i\hat{\theta}_{(1,m)}} e^{i H t} c_{(1,m^\prime) }e^{i\hat{\theta}_{(1,m^\prime)}} | 0 \rangle.
\nonumber \\  \label{eq:edgeSSF_2}
\end{eqnarray}
Using the facts that $e^{i\hat{\theta}_{(1,m)}}$ commutes with $H$ and that 
for any two states of $c$-Majorana fermions the matrix element of $e^{i \hat{\theta}_{(1,m)}} e^{i \hat{\theta}_{(1,m^\prime)}}$ is zero for $m \neq m^\prime$~\cite{Mizoguchi2019},
we find that only the terms with $m = m^\prime$ have a finite contribution in the summation of (\ref{eq:edgeSSF_2}).
Then, we obtain
\begin{eqnarray}
\chi_{\rm edge} (\omega)= &\frac{1}{L_y} \sum_{m}
\int_{-\infty}^{\infty} dt e^{- i \omega t} \langle 0 |  c_{(1,m)}  e^{i H t} c_{(1,m) } | 0 \rangle,\nonumber \\ 
= & \int_{-\infty}^{\infty} dt e^{- i \omega t} \langle 0 |  c_{(1,1)}  e^{i H t} c_{(1,1) } | 0 \rangle,\nonumber \\
 \label{eq:edgeSSF_3}
\end{eqnarray}
where we have set $H |0 \rangle = 0$ for simplicity, and have used the translational symmetry along the vertical direction to obtain the second line. 

To proceed, we employ a spectral representation:
\begin{eqnarray}
\chi_{\rm edge} (\omega)=& \sum_{\lambda} \int_{-\infty}^{\infty} dt e^{- i \omega t} 
e^{iE_{\lambda} t} 
\langle 0 |  c_{(1,1)} |\lambda \rangle \langle \lambda | c_{(1,1) } | 0 \rangle,\nonumber \\
= & 2 \pi  \sum_{\lambda} \delta(\omega - E_\lambda)
\langle 0 |  c_{(1,1)} |\lambda \rangle \langle \lambda | c_{(1,1) } | 0 \rangle,\nonumber \\
 \label{eq:edgeSSF_4}
\end{eqnarray}
where $|\lambda\rangle $ denotes the eigenstate of $H$ that satisfies $H|\lambda\rangle = E_{\lambda } |\lambda\rangle$.
To calculate the expectation value $\langle \lambda | c_{(1,1)} | 0 \rangle$ in Eq. (\ref{eq:edgeSSF_4}),
we expand $c_{(1,1)}$ in terms of $\gamma^{\rm L} (k_y)$.
To do this, let us first recall that
\begin{eqnarray}
c_{(1,1)} = &\alpha_{(1,1)} + \alpha^{\dagger}_{(1,1)}  \nonumber \\
= & \frac{1}{\sqrt{L_y}} \sum_{k_y} e^{- i k_y} \left(\alpha_{1,k_y}  +\alpha^\dagger_{1,-k_y} \right).
\end{eqnarray}
We expand $\alpha_{1,k_y}$ as 
\begin{eqnarray}
\alpha_{1,k_y} = W_{k_y} \gamma^{\rm L} (k_y)  + \cdots.
\end{eqnarray}
Note that the anti-commutator $\{ \alpha_{1,k_y} , [\gamma^{\rm L} (k_y)]^\dagger \}$ is equal to $W^\ast_{k_y}$,
since $\{  \gamma^{\rm L} (k_y) , [ \gamma^{\rm L} (k_y) ]^\dagger\} = 1$.
On the other hand, from Eq. (\ref{eq:leftedge}), we have $\{ \gamma^{\rm L} (k_y) , \alpha^\dagger_{1,k_y} \} = 
\left[\{ \alpha_{1,k_y} , [\gamma^{\rm L} (k_y)]^\dagger \} \right]^\ast
= u_{1,k_y}$,
since $\{  \alpha_{1,k_y} ,\alpha_{1,k_y}^\dagger\} = 1$.
Combining these, we obtain $W_{k_y} = u_{1,k_y}^\ast$.
Therefore, we have 
\begin{eqnarray}
{\scriptstyle 
 c_{(1,1)} = 
 \frac{1}{\sqrt{L_y}} \sum_{k_y} e^{- i k_y} \left\{u^\ast_{1, k_y} \gamma^{\rm L} (k_y)  
+ u_{1,-k_y} \left[\gamma^{\rm L} (-k_y)\right]^\dagger   + \cdots \right\}. }
\nonumber \\
\label{eq:c_expand}
\end{eqnarray}
The rest except for the first two terms in the summand in the right-hand side of (\ref{eq:c_expand}) 
do not contribute to the edge susceptibility because the frequency $\omega$ is assumed to be smaller than the bulk band gap.

Using Eq.~(\ref{eq:c_expand}) and noting $\gamma^{\rm L} (k_y) |0\rangle = 0$ $\{[\gamma^{\rm L} (k_y)]^\dagger |0\rangle = 0\}$
for $\varepsilon^{\rm L} (k_y) > 0$ [$\varepsilon^{\rm L} (k_y) < 0$],
we have 
\begin{widetext}
\begin{eqnarray}
\chi_{\rm edge} (\omega)
= & \frac{2\pi}{L_y} \sum_{k_y} \sum_{k_y^\prime} \sum_{\lambda} \delta(\omega - E_\lambda)
e^{i(k_y^\prime - k_y)} u^{\ast}_{1, k_y} u_{1,k_y^\prime}
\langle 0 | \gamma^{\rm L} (k_y) |\lambda \rangle \langle \lambda | \left[ \gamma^{\rm L} (k_y^\prime) \right]^\dagger | 0 \rangle.
 \label{eq:edgeSSF_5}
\end{eqnarray}
\end{widetext}
The matrix element $\langle 0 | \gamma^{\rm L} (k_y) | \lambda \rangle$ 
becomes finite when $ | \lambda \rangle = \left[\gamma^{\rm L} (k_y)\right]^\dagger |0 \rangle$ and $\varepsilon^{\rm L} (k_y) > 0$,
or $ | \lambda \rangle = \gamma^{\rm L} (k_y)  |0 \rangle$ and $\varepsilon^{\rm L} (k_y) < 0$.
Therefore, we have
\begin{eqnarray}
\chi_{\rm edge} (\omega)
= & \frac{2\pi}{L_y}  \sum_{k_y} |u_{1, k_y} |^2 \delta(\omega - E^{\rm L} (k_y)) \nonumber \\
= & \int d \epsilon \left(\frac{d k_y}{d \epsilon}\right) |u_{1, k_y} |^2  \delta(\omega - \epsilon) \nonumber \\
= & |u_{1, \tilde{k}_\omega} |^2 \left[v^{\rm L} (\tilde{k}_\omega) \right]^{-1}
 \label{eq:edgeSSF_6}
\end{eqnarray}
where we have introduced $\tilde{k}_\omega$ such that $ E^{\rm L} (\tilde{k}_\omega)= \omega$.

From Eq. (\ref{eq:edgeSSF_6}), we find that $\chi_{\rm edge} (\omega)$
is proportional to the inverse of the group velocity,
meaning that the sign of the $\chi_{\rm edge} (\omega)$ is determined by that of the group velocity.
Hence, by measuring the $\omega$ dependence of $\chi_{\rm edge} (\omega)$,
one can determine whether the edge mode is chiral or not.
To be concrete, if the edge mode is chiral, then the sign is always positive or negative, 
whereas if it is not chiral, then the sign change occurs on varying $\omega$.

\section{Summary and discussion \label{sec:summary}}
In summary, we have treated the propagation of the edge magnetization 
due to the chiral edge modes and the edge dynamical susceptibility 
in the Kitaev honeycomb model with the three-spin interaction and the Kitaev 
triangle-honeycomb model. These quantities have been shown to refelect the chirality of 
the edge modes, whose emergence is a consequence of the topological nature of 
these models, i.e., the nontrivial Chern number and the bulk-edge correspondence. 
By relying on the exact solvability of these models, we have first mapped them to 
the free fermion models, and then applied the transfer matrix method.  
Further, by using the resulting edge wave functions and their dispersion relations, 
we have calculated (i) the time evolution of the edge magnetization and 
(ii) the edge dynamical susceptibility. 
For the former (i), we have demonstrated that 
the wave packet of the edge magnetization due to the chiral edge mode 
which is excited by an external perturbation indeed moves in only one of 
the two directions of the edge with the group velocity. For the latter (ii), 
we have shown that the edge dynamical susceptibility is proportional to 
the inverse of the group velocity with the positive weight. This property enables us
to determine whether an edge mode is chiral or not. In fact, if an edge mode is chiral, 
then the sign of the edge dynamical susceptibility does not change 
through the whole range of the frequency; otherwise, it takes both of plus and minus signs. 

Our method is also applicable to generic topological insulators and superconductors, 
whose concrete examples are the Haldane model~\cite{Haldane1988} and the spinless superconductor~\cite{Sato2016},  
which were already treated in the text. In particular, the oriented propagation of 
the excess charge can be expected to be experimentally detected in a chiral superconductor.  

Before closing this paper, we address the experimental observation of the oriented propagation of the edge magnetization.
As we have seen, the propagation can be expected to occur
by exciting the chiral edge spinons by shining a laser pulse to the Kitaev magnets under the magnetic field in [111] direction. 
Then, we expect that the flow of the magnetization can be detected by using the sophisticated techniques developed in the field of spintronics~\cite{Bakun1984,Saitoh2006,Kimura2007,Maekawa2013}. 
In particular, the propagation of the edge magnetization can be expected 
to be stable against disorder that is inevitable in experimental situations 
because the chiral edge modes are known to be robust against perturbations~\cite{Schulz-Baldes2000,Kellendonk2002,Elbau2002}. 

\acknowledgements
We wish to thank H. Katsura and M. Udagawa for fruitful discussion.
T. M. is supported by the JSPS KAKENHI (Grant No. JP17H06138), MEXT, Japan. 

\appendix
\begin{widetext}
\section{Transfer matrix method for the edge modes \label{sec:edgemode_method}}
In this Appendix, we present the details of the transfer matrix method 
by which we obtain the dispersion relation 
of the left edge mode in Sec.~\ref{sec:chiraledge_transfer}. 

To begin with, we recall that the BdG equation of Eq. (\ref{eq:BDG}) is expressed 
by using the transfer matrix $\hat{T}(k_y)$ as Eq. (\ref{eq:TM_eq}).
Let us consider the eigenvalue problem for $\hat{T}(k_y)$.
In the following, we abbreviate generally a function of $f(k_y)$ of $k_y$ to $f$ by dropping the argument $k_y$. 

From (\ref{eq:TM_eq}) and (\ref{eq:TM_def}), the eigenvalue equation for $\hat{T}=\hat{T}(k_y)$ can be written as
\begin{eqnarray}
\left(
\begin{array}{c}
\bm{\varphi}_{\ell+1} \\ 
\bm{\varphi}_{\ell} 
\end{array} 
\right)=
\left(
\begin{array}{cc}
\hat{A}^{-1}[\varepsilon^{\mathrm{L}}-\hat{B}] & -\hat{A}^{-1} \hat{A}^\dagger \\ 
\hat{I}_{2} & 0 \\
\end{array}
\right)
\left(
\begin{array}{c}
\bm{\varphi}_{\ell} \\ 
\bm{\varphi}_{\ell-1} \end{array}
\right)
=\lambda 
\left(
\begin{array}{c}
\bm{\varphi}_{\ell} \\ 
\bm{\varphi}_{\ell-1} 
\end{array} \right), \label{eq:trans_eigen}
\end{eqnarray}
with an eigenvalue $\lambda$. 
Although the size of $\hat{A}$ and $\hat{B}$ is $2 \times 2$ in the present problem, 
the formalism of the transfer matrix is applicable 
to the case with any size of $n \times n$ ($n=1,2, \cdots$).
Therefore, in what follows, we consider the case for generic $n$, 
whose eigenvalue equation can be obtained by replacing $\hat{I}_2$ with $\hat{I}_n$ in Eq. (\ref{eq:trans_eigen}). 
We will get back to the case of $n=2$ after we show the generic formulation.

From Eq. (\ref{eq:trans_eigen}), one has 
\begin{eqnarray}
\bm{\varphi}_{\ell} =\lambda\bm{\varphi}_{\ell-1},
\end{eqnarray}
and 
\begin{equation}
\hat{A}^{-1}(\varepsilon^{\mathrm{L}} - \hat{B}) \lambda\bm{\varphi}_{\ell-1}- \hat{A}^{-1}\hat{A}^\dagger\bm{\varphi}_{\ell-1}
=\lambda^2\bm{\varphi}_{\ell-1}.
\label{EigenEq}
\end{equation}
When $ \lambda \neq 0$, this can be written as     
\begin{eqnarray}
\lambda \hat{A} \bm{\varphi}_{\ell-1}+\lambda^{-1} \hat{A}^\dagger \bm{\varphi}_{\ell-1}+(\hat{B}-\varepsilon^{\mathrm{L}})\bm{\varphi}_{\ell-1}=0.
\end{eqnarray}
For a nontrivial solution $\bm{\varphi}_{\ell-1}$, the two parameters, 
$\lambda$ and $\varepsilon^{\mathrm{L}}$, must satisfy 
\begin{eqnarray}
{\rm det}[\lambda \hat{A}+\lambda^{-1} \hat{A}^\dagger +(\hat{B}-\varepsilon^{\mathrm{L}})]=0.
\end{eqnarray}
Further, using the fact that if the matrix $M$ satisfies ${\rm det}M=0$, then ${\rm det}M^\dagger=0$, 
we have 
\begin{eqnarray}
{\rm det}[\lambda^\ast \hat{A}^\dagger+{(\lambda^\ast)}^{-1} \hat{A} +(\hat{B}-\varepsilon^{\mathrm{L}})]=0,
\end{eqnarray}
where we have used that $\hat{B}$ is Hermitian, and that $\varepsilon^{\mathrm{L}}$ is real. 
This implies the following:  
If $\lambda$ is a solution of the above equation for a given real $\varepsilon^{\mathrm{L}}$, 
then $1/\lambda^\ast$ is a solution, too~\cite{Molinari1997}. 
Thus the equation has a set of the solutions, 
\begin{eqnarray}
\lambda_1, 1/\lambda_1^\ast,\lambda_2,1/\lambda_2^\ast, \ldots, \lambda_n,1/\lambda_n^\ast.
\end{eqnarray}
Here, we assume that the eigenvalues, $\lambda_j$, 
satisfy $0<|\lambda_j|<1$ for 
$\ell=1,2,\ldots,n$. In the present approach, we need this assumption in order to obtain 
a decay solution with the open boundary condition at the edge. 
For the corresponding $n$-component eigenvector which satisfies (\ref{EigenEq}) 
with the eigenvalue $\lambda_j$, we write $\bm{\chi}^{(j)}$. 

Then, the $2n$-component vector, 
\begin{eqnarray}
\left(
\begin{array}{c}
\lambda_{j} \bm{\chi}^{(j)} \\ 
\bm{\chi}^{(j)} 
\end{array}
\right), 
\end{eqnarray}
is an eigenvector of the transfer matrix. 

In order to find the solution 
which satisfies the open boundary condition at the edge, we set 
\begin{eqnarray}
\left(\begin{array}{c}\bm{\varphi}_\ell \\ \bm{\varphi}_{\ell-1} \end{array} \right)=\sum_{j=1}^n c_{j} 
(\lambda_{j})^{\ell-1}
\left(\begin{array}{c} \lambda_{j} \bm{\chi}^{(j)} \\ \bm{\chi}^{(j)} \end{array}\right). \label{eq:chi_def}
\end{eqnarray}
The coefficients $c_\ell$ are determined by 
\begin{eqnarray}
\sum_{j=1}^n c_j \bm{\chi}^{(j)}=0 \label{eq:chi_c}
\end{eqnarray}
because of the open boundary condition $\bm{\varphi}_0=0$. This implies 
\begin{equation}
{\rm det}[\bm{\chi}^{(1)},\bm{\chi}^{(2)},\ldots,\bm{\chi}^{(n)}]=0. 
\label{detEq}
\end{equation}
Since all the eigenvalues $\lambda_j$ and all the vectors $\bm{\chi}^{(j)}$ are determined by 
the energy $\varepsilon^{\mathrm{L}}$, this equation determines the energy spectrum $\varepsilon^{\mathrm{L}}$ of the edge mode. 
But it is very difficult to determine the energy $\varepsilon^{\mathrm{L}}$ by the above equation (\ref{detEq})
for general $n$. 

Now, let us consider the case of $n=2$, which corresponds to our original problem for the Kitaev model. 
In this case, one clearly has 
\begin{eqnarray}
{\rm det}[\bm{\chi}^{(1)},\bm{\chi}^{(2)}]=0.
\end{eqnarray}
This implies that the two vectors are the same, i.e, $\bm{\chi}^{(1)}={\rm(const.)}\bm{\chi}^{(2)}$. 
In addition to this, if $\lambda_1 = \lambda_2$, then $\varphi_\ell = 0$
for all $\ell$ from (\ref{eq:chi_def}) and (\ref{eq:chi_c}). Therefore, in this case, the single vector $\bm{\chi}^{(1)}$ has two different eigenvalues, 
$\lambda_1$ and $\lambda_2$, 
i.e., $\lambda_1\ne \lambda_2$. 
Then, the corresponding equations for (\ref{EigenEq}) 
become 
\begin{eqnarray}
\hat{A}^{-1}(\varepsilon^{\mathrm{L}}-\hat{B})\lambda_1\bm{\varphi}^1
-\hat{A}^{-1} \hat{A}^\dagger \bm{\varphi}^1=\lambda_1^2\bm{\varphi}^1 \label{eq:tm_app_1}
\end{eqnarray}
and
\begin{eqnarray}
\hat{A}^{-1}(\varepsilon^{\mathrm{L}}- \hat{B})\lambda_2\bm{\varphi}^2
-\hat{A}^{-1} \hat{A}^\dagger\bm{\varphi} ^2=\lambda_2^2\bm{\varphi}^2. \label{eq:tm_app_2}
\end{eqnarray}
Subtracting (\ref{eq:tm_app_2}) from (\ref{eq:tm_app_1})
and using $\bm{\varphi}^2=\bm{\varphi}^1$, one has 
\begin{eqnarray}
\hat{A}^{-1}(\varepsilon^{\mathrm{L}}-\hat{B})(\lambda_1-\lambda_2)\bm{\varphi}^1
=(\lambda_1^2-\lambda_2^2) \bm{\varphi}^1.
\end{eqnarray}
This implies that $\bm{\varphi}^1$ is the eigenvector of the matrix $\hat{A}^{-1}(\hat{B}-\varepsilon^{\mathrm{L}})$. 
Then, clearly, $\bm{\varphi}^1$ is the eigenvector of $\hat{A}^{-1}\hat{A}^\dagger$ as well. 

Since the matrix $\hat{A}^{-1}\hat{A}^\dagger$ is independent of the energy $\varepsilon^{\mathrm{L}}$, 
one can obtain an eigenvector $u$ of $\hat{A}^{-1}\hat{A}^\dagger$ with the eigenvalue $\bar{\mu}$ which satisfies 
the decay condition. Then,  
\textit{we can choose the energy eigenvalue $\varepsilon^{\mathrm{L}}$ so that $u$ becomes the eigenvector of $\hat{A}^{-1}(\hat{B}-\varepsilon^{\mathrm{L}})$}. 
More precisely, $\varepsilon^{\mathrm{L}}$ can be chosen so that the ratios of the first and the second components of the vectors 
in both sides of the eigenvalue equation $\hat{A}^{-1}(\hat{B}-\varepsilon^{\mathrm{L}})u=\bar{\eta} u$ 
with an eigenvalue $\bar{\eta}$ coincide with each other.  
Indeed, the dispersion relation of (\ref{eq:edge_disp}) is determined in this way. 
In consequence, we have 
\begin{eqnarray}
\hat{A}^{-1}(\hat{B}-\varepsilon^{\mathrm{L}}) u=\bar{\eta}u
\end{eqnarray}
with the eigenvalue $\bar{\eta}$. 
Combining this, $\hat{A}^{-1}\hat{A}^\dagger u= \bar{\mu} u$ 
and Eq.~(\ref{EigenEq}) with $\bm{\varphi}_{\ell-1}=u$, we have 
\begin{eqnarray}
-\lambda\bar{\eta}-\bar{\mu}=\lambda^2. 
\end{eqnarray}
The two eigenvalues of the transfer matrix are given by 
\begin{eqnarray}
\lambda_\pm = \frac{-\bar{\eta}\pm\sqrt{\bar{\eta}^2-4\bar{\mu}}}{2}=
-\frac{\bar{\eta}}{2}\pm \sqrt{\left(\frac{\bar{\eta}}{2}\right)^2-\bar{\mu}}.
\end{eqnarray}
Consequently, the wave function of the edge mode is given by 
\begin{eqnarray}
\bm{\varphi}_\ell=(\lambda_+^\ell -\lambda_-^\ell) u,
\end{eqnarray}
which satisfies the boundary conditions when 
\begin{eqnarray}
|\lambda_\pm|<1. \label{eq:lambda_cond}
\end{eqnarray}
The condition (\ref{eq:lambda_cond}) is the criterion of 
the existence of the edge mode at given $k_y$ (see the area shaded in yellow in Fig.~\ref{Fig2}).

\section{Chiral edge mode in the Kitaev triangle-honeycomb model \label{sec:trihoney}}
In this Appendix, we explain how to construct the solution for the edge modes of the Kitaev triangle-honeycomb model.
To begin with, we expand the $i$-th edge mode as 
\begin{equation}
\gamma^{(i)}_{k_y} = \sum_{\ell = 1}^{L_x} \sum_{s = A,B,C} u^{(i)}_{\ell, k_y, s} \alpha_{\ell, k_y, s} + v^{(i)}_{\ell, k_y, s} \alpha^\dagger_{\ell,-k_y,s},
\end{equation}
where the coefficients $u^{(i)}_{\ell,k_y,\eta}$ and $v^{(i)}_{\ell,k_y,\eta}$ satisfy the BdG equation:  
\begin{equation}
\sum_{\ell^\prime = 1}^{L_x}
\left\{ [\hat{h}  (k_y)]_{\ell, \ell^\prime}  
-\varepsilon^{\mathrm{L},(i)}(k_y) \hat{I}_6 \delta_{\ell,\ell^\prime} \right \}
\left(
\begin{array}{c}
u^{(i)}_{\ell^\prime,k_y,A} \\
u^{(i)}_{\ell^\prime,k_y,B} \\
u^{(i)}_{\ell^\prime,k_y,C} \\
v^{(i)}_{\ell^\prime,k_y,A} \\
v^{(i)}_{\ell^\prime,k_y,B} \\
v^{(i)}_{\ell^\prime,k_y,C} \\
\end{array}
\right)
 = 0, \label{eq:trihoney_eigen}
\end{equation}
for $\ell = 1, \cdots L_x$. 
To solve Eq. (\ref{eq:trihoney_eigen}), we define
\begin{eqnarray}
\xi^{(i)}_{\ell,k_y, s} :=u_{\ell,k_y,s} - v_{\ell,k_y,s},
\end{eqnarray}
and 
\begin{eqnarray}
\zeta^{(i)}_{\ell,k_y, s} :=u_{\ell,k_y,s} + v_{\ell,k_y,s}
\end{eqnarray}
for $s = A,B,C$.
Then, among six equations of (\ref{eq:trihoney_eigen}), 
the four equations contain only the variables at the site $\ell$.
These are given by
\begin{subequations}
\begin{equation}
 J^\prime \zeta^{(i)}_{\ell,k_y, B}  -\varepsilon^{L,(i)} (k_y) \xi^{(i)}_{\ell,k_y, B}  \label{eq:rec_1}
 = J \zeta^{(i)}_{\ell,k_y, C} - iJ  \xi^{(i)}_{\ell,k_y, C},
\end{equation}
\begin{equation}
 J^\prime \xi^{(i)}_{\ell,k_y, B}-\varepsilon^{L,(i)} (k_y) \zeta^{(i)}_{\ell,k_y, B}  =  J  \xi^{(i)}_{\ell,k_y, A}  - iJ \zeta^{(i)}_{\ell,k_y, A}, \label{eq:rec_2}
\end{equation}
\begin{equation}
J \zeta^{(i)}_{\ell,k_y,A} -\varepsilon^{L,(i)}(k_y)  \xi^{(i)}_{\ell,k_y,A} = - J^\prime e^{-ik_y} \zeta^{(i)}_{\ell,k_y,C} + J \zeta^{(i)}_{\ell,k_y,B}, \label{eq:rec_4}
\end{equation}
\begin{equation}
J \xi^{(i)}_{\ell,k_y,C} -\varepsilon^{L,(i)}(k_y) \zeta^{(i)}_{\ell,k_y,C}= J \xi^{(i)}_{\ell ,k_y,B} - J^\prime e^{ik_y} \xi^{(i)}_{\ell,k_y,A}.  \label{eq:rec_5}
\end{equation}
\end{subequations}
The rest of the two equations are
\begin{subequations}
\begin{equation}
J \xi^{(i)}_{\ell,k_y,A} -\varepsilon^{L,(i)}(k_y)  \zeta^{(i)}_{\ell,k_y,A}=  iJ \zeta^{(i)}_{\ell,k_y,B} + J^\prime \xi^{(i)}_{\ell-1,k_y,C}, \label{eq:rec_3}
\end{equation}
\begin{equation}
 J \zeta^{(i)}_{\ell,k_y,C}  -\varepsilon^{L,(i)}(k_y)   \xi^{(i)}_{\ell,k_y,C}=  i J \xi^{(i)}_{\ell,k_y,B} + J^\prime \zeta^{(i)}_{\ell+1,k_y,A}. \label{eq:rec_6}
\end{equation}
\end{subequations}

We write 
\begin{eqnarray}
\bm{\Phi}_\ell^{(i)}(k_y)=\left(\begin{array}{c}
\zeta_{\ell,k_y,A}^{(i)}\\
\xi_{\ell,k_y,A}^{(i)}\\
\zeta_{\ell,k_y,B}^{(i)}\\
\xi_{\ell,k_y,B}^{(i)}\\
\zeta_{\ell,k_y,C}^{(i)}\\
\xi_{\ell,k_y,C}^{(i)}
\end{array}
\right),
\end{eqnarray}
\begin{eqnarray}
\hat{A}(k_y)=\left(
\begin{array}{cccc}
0 & 0 & \cdots & 0\\
\vdots & \vdots & \ddots & \vdots \\
0 & 0 & \cdots & 0\\ 
J^\prime & 0 & \cdots & 0\\
\end{array}\right), 
\end{eqnarray}
and 
\begin{eqnarray}
\hat{B}(k_y)=
\left(\begin{array}{cccccc}
0 & J & -iJ & 0 & 0 & 0\\
J & 0 & -J & 0 & J^\prime e^{-ik_y} & 0\\
iJ & -J & 0 & J^\prime & 0 & 0\\
0 & 0 & J^\prime & 0 & -J & iJ \\
0 & J^\prime e^{ik_y} & 0 & -J &  0 & J\\ 
0 & 0 & 0 & -iJ & J & 0\\
\end{array}
\right). 
\end{eqnarray}
Then, the equation corresponding to Eq.~(\ref{eq:eigen}) is given by 
\begin{eqnarray}
\hat{A}(k_y)\bm{\Phi}_{\ell+1}^{(i)}+\hat{A}^\dagger(k_y)\bm{\Phi}_{\ell-1}^{(i)}+\hat{B}(k_y)\bm{\Phi}_\ell^{(i)}
=\varepsilon^{L,(i)}\bm{\Phi}_\ell^{(i)}. \label{eq:trihoney_eq_1}
\end{eqnarray}
Although the matrix $\hat{A}(k_y)$ is not invertible~\cite{Dwivedi2016}, we can find the left edge solutions 
which satisfy $\xi_{\ell,k_y,C}^{(i)}=0$ for all $\ell$. 
Actually, when choosing $\xi_{\ell,k_y,C}^{(i)}=0$, the rest of the five components satisfy    
\begin{eqnarray}
\left(
\begin{array}{cccccc}
-\varepsilon^{L,(i)} & J & -iJ & 0 & 0 \\
J & -\varepsilon^{L,(i)} & -J & 0 & J^\prime e^{-ik_y}\\
iJ & -J & -\varepsilon^{L,(i)} & J^\prime & 0\\
0 & 0 & J^\prime & -\varepsilon^{L,(i)} & -J\\
0 & J^\prime e^{ik_y} & 0 & -J &  -\varepsilon^{L,(i)}\\
\end{array}
\right)
\left(
\begin{array}{c}
\zeta_{\ell,k_y,A}^{(i)}\\
\xi_{\ell,k_y,A}^{(i)}\\
\zeta_{\ell,k_y,B}^{(i)}\\
\xi_{\ell,k_y,B}^{(i)}\\
\zeta_{\ell,k_y,C}^{(i)}\\
\end{array}
\right) = 0. \label{eq:trihoney_eq_2}
\end{eqnarray}
Since this eigenvalue equation must have a nontrivial solution, the determinant of the matrix 
must be vanishing. As a result, the energy eigenvalue $\varepsilon^{L,(i)}$ can be determined by 
\begin{eqnarray}
[\varepsilon^{\rm L,(i)}]^5 -(4J^2 -2J^{\prime 2})[\varepsilon^{\rm L,(i)}]^3 + (3J^4 + 2 J^2 J^{\prime 2}  + J^{\prime 4} -2 J^2 J^{\prime 2} \cos k_y) \varepsilon^{\rm L,(i)} + 2J^3 J^{\prime 2} \sin k_y = 0. \label{eq:trihoney_edge_disp}
\end{eqnarray}
Since the above matrix is hermitian, this algebraic equation has five real solutions 
which are already found in Fig.~\ref{Fig7} as edge modes. 
Further, the nontrivial solution of (\ref{eq:trihoney_eq_2}) is determined except for the normalization. 
Combining this with the equation for the sixth component of (\ref{eq:trihoney_eq_1}), we obtain the recursion equation 
$\zeta_{\ell+1,k_y,A}^{(i)}=\lambda(k_y)\zeta_{\ell,k_y,A}^{(i)}$ with 
\begin{eqnarray}
\lambda(k_y)=  \frac{e^{ik_y} \left\{-[\varepsilon^{\rm L,(i)}]^4+i [\varepsilon^{\rm L,(i)}]^3 
 J+[\varepsilon^{\rm L,(i)}]^2 \left(3 J^2+J^{\prime 2}\right)-i \varepsilon^{\rm L,(i)} J \left(3 J^2-J^{\prime 2} e^{ik_y}\right)-J^2 J^{\prime 2} 
 \left(1+e^{ik_y}\right)\right\}}{J^{\prime 2} \left\{e^{ik_y} \left(-[\varepsilon^{\rm L,(i)}]^2+i \varepsilon^{\rm L,(i)} J+J^{\prime 2}\right)+i J (\varepsilon^{\rm L,(i)}+i J)\right\} }.
\end{eqnarray}
Clearly, the left-edge solutions must satisfy the decay condition $|\lambda(k_y)|<1$ which determines 
the permissible region of $k_y$ for given $J$ and $J^\prime$. 
Figure~\ref{Fig8} shows the region for the edge modes across the zero energy. 
As seen in Fig.~\ref{Fig7}, the permissible wave number $k_y$ for the edge modes is restricted to the neighborhood of $\pi$ 
in the topological phase, whereas, in the nontopological phase, the permissible wave number can take 
any value in the whole range. 
Similarly, the right edge solutions 
satisfy $\zeta_{\ell,k_y,A}^{(i)}=0$ for all $\ell$, and we can obtain them in the same way. 
Although we cannot solve the algebraic equation (\ref{eq:trihoney_edge_disp}) analytically,  
we believe that the useful information about the chiral edge modes has been able to be obtained. 
\begin{figure}[!htb]
\begin{center}
\includegraphics[width= 0.4\linewidth]{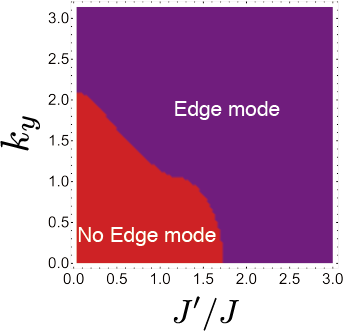}
\caption{A parameter space of $J^\prime/J$ and $k_y$.
A region colored in red does not have the solution of the edge mode, i.e., $|\lambda(k_y)| > 1$,
and a region colored in purple has the solution of the edge mode, i.e., $|\lambda(k_y)| < 1$. }
\label{Fig8}
\end{center}
\end{figure}
\end{widetext}

\end{document}